\documentclass{rstransa}
\usepackage{color} 
\definecolor{cyanp}{rgb}{0.11, 0.59, 0.73}
\usepackage{graphicx}

\newcommand{\be}{\begin{equation}}
\newcommand{\ee}{\end{equation}}
\newcommand{\bea}{\begin{eqnarray}}
\newcommand{\eea}{\end{eqnarray}}
\newcommand{\lb}{\label}

\newcommand{\bF}{{\bf f}}

\newcommand{\bv}{{\bf v}}
\newcommand{\bu}{{\bf u}}

\newcommand{\bk}{{\bf k}}

\newcommand{\bx}{{\bf x}}
\newcommand{\br}{{\bf r}}

\newcommand{\bX}{{\bf X}}

\newcommand{\grad}{{\mbox{\boldmath $\nabla$}}}
\newcommand{\bdot}{{\mbox{\boldmath $\cdot$}}}

\def\Re{\textrm{Re}}

\begin{document}

%

\title{Beyond chaos: fluctuations, anomalies and spontaneous stochasticity in fluid turbulence}

\author{
Gregory L. Eyink$^{1}$ and Nigel Goldenfeld$^{2}$}

\address{$^{1}$Department of Applied Mathematics \& Statistics and Department of Physics \& Astronomy, 
The Johns Hopkins University, Baltimore, Maryland, 21218, USA\\ 
$^{2}$Department of Physics, University of California San Diego, 9500 Gilman Drive, La Jolla, CA 92093, USA}

\subject{xxxxx, xxxxx, xxxx}

\keywords{xxxx, xxxx, xxxx}

\corres{Gregory Eyink and Nigel Goldenfeld\\
\email{eyink@jhu.edu and nigelg@ucsd.edu}}

\begin{abstract}
In this perspective, we consider the development of statistical hydrodynamics, focusing on the way in which the intrinsic stochasticity of turbulent phenomena was identified and is being explored.  A major purpose of our discussion is to bring out the role of anomalies in turbulent phenomena, in ways that are not usually done, and to emphasize how the description of turbulent phenomena requires delicate considerations of asymptotic limits.
The scope of our narrative includes selected historical aspects that are not usually emphasized, primarily due to G.I. Taylor, as well as discussions of certain aspects of the laminar-turbulent transition, the behaviour of turbulent drag at intermediate Reynolds numbers, and the statistics of fully-developed turbulence that exhibit spontaneous stochasticity.  

\bigskip
\noindent
This article is part of the themed issue ``Frontiers of Turbulence and Statistical Physics Meet''. 
\end{abstract}


\begin{fmtext}
\end{fmtext} 

\maketitle 

\section{Introduction} 

The explanatory power of statistical mechanics stems from its capability to extract universal features from physical phenomena. In the case of turbulence in simple fluids, there are two specific regimes where universal phenomena might occur: transitional phenomena and fully-developed turbulence.  In this article, we will present some commentary and interpretation of recent advances made in both these areas, from the perspective of statistical mechanics.  We will see that subtle issues of asymptotics, renormalization group, and scaling laws are central to understanding what we mean by \lq\lq an explanation of turbulence \rq\rq, and we will try to resolve some of the repeated misunderstandings that are encountered in the literature.  Our focus will be on recent advances in understanding the laminar-turbulent transition, the behaviour of dissipation in fluids at Reynolds numbers above the onset of turbulence, and the nature of the fluctuations encountered for asymptotically large Reynolds numbers.  We will emphasize that fully-developed turbulence cannot be thought of as merely chaotic, but that the fluctuations and divergence of Lagrangian trajectories are in a sense 
``super-chaotic'', with Lyapunov exponents diverging at asymptotically large Reynolds 
numbers and leading to fundamentally new phenomena. 

Our perspective is that whilst it is a truism that turbulence is a stochastic phenomenon, the precise characterization of that stochasticity has remained elusive.  In particular, we argue that until very recently, the nature of turbulent statistics was a matter of assumption, and only with two advances, has it become clear how to predict the intrinsic randomness exhibited by turbulence.  The two settings where this has been the case are: (1) the laminar-turbulent transition in wall-bounded shear and pressure-driven flows, and (2) the statistics of passive scalar advection, and by extension presumably other high Reynolds number flows. For (1), it has become clear from both theory and experiment \cite{sreenivasan1986transition,crutchfield1988attractors,pomeau,manneville2016transition,
barkley2016theoretical,goldenfeld2017statistical,
goldenfeld2017turbulence,eckhardt2018transition,
mukund2018critical,avila2023transition,hof2008repeller,goldenfeld2010extreme,
avila2011onset,barkley2011simplifying,
sipos2011directed,barkley2012pipe,song2014deterministic,barkley2015rise,
hof_dp_1d_couette, ppmodel,chantry2017universal,
budanur2020upper,gome2020statistical,wang2022stochastic,
klotz2022phase,avila2023transition,hof2023directed,lemoult2024directed} that these nominally deterministic Navier-Stokes flows become turbulent through a non-equilibrium continuous phase transition in the universality class of directed percolation \cite{hinrichsen2000non}, even though such flows show sub-critical transitions.  For (2), it has emerged from exactly soluble models that the statistics of Lagrangian trajectories in fully-developed turbulence may exhibit so-called spontaneous stochasticity, with fluctuations that are stronger and distinct from chaos \cite{lorenz1969predictability,bernard1998slow,gawedzki2000phase,E2000generalized,
lejan2002integration,lejan2004flows,palmer2014real,eyink2015spontaneous,
mailybaev2016spontaneously,thalabard2020butterfly,drivas2017Alagrangian,drivas2024statistical,bandak2024spontaneous}. These predictions are supported indirectly by experiment and confirmed by numerical simulations, but the immediate effects have not yet been observed in laboratory flows. 

A central motif of our narrative will be that seemingly negligible physical variables may nevertheless not be discarded, because they lead to qualitatively new phenomena, whilst at the same time not modifying the quantitative values of observables, and thus generating universal phenomena.  This we discuss in relation to the dissipative anomaly in turbulence, which is perhaps the most profound expression of the nature of turbulent stochasticity, and speculate that such anomalies are present in wave turbulence too.

This article is organized as follows.  In Section \ref{sec:dawn}, we present a brief historical discussion of the work of G.I. Taylor that introduced statistical reasoning into turbulence.  Taylor was concerned very early on by the nature of the laminar-turbulent transition, and in Section \ref{sec:trans} we briefly review recent work on pipe transitional flows, emphasizing how a surprising statistical result emerges from experiments and theoretical work.  Above the laminar-turbulent transition, especially in pipes, the dissipation (or drag) behaves in a surprisingly non-monotonic way, that we show in Section \ref{sec:diss} is a manifestation of two types of dissipative anomaly.  
In Section \ref{sec:spontstoch}, we introduce the notion of spontaneous stochasticity, and emphasize how it may arise in deterministic Navier-Stokes fluids at high Reynolds numbers.  Finally, in Section \ref{sec:disc} we conclude with a synthesis of the key points that are common to all the phenomena discussed.

\section{The dawn of statistical reasoning in turbulence}
\label{sec:dawn}
For over a hundred years, fully-developed turbulence has been conceptualized as a statistical process.  Following the work of Osborne Reynolds introducing the eponymous decomposition of velocity into mean and fluctuating components \cite{reynolds1895iv}, G.I. Taylor was probably the first person to develop sophisticated statistical assumptions for turbulence, in order to understand how it transported heat \cite{taylor1915eddy,taylor1922diffusion} (see also his riveting historical recollections on the vicissitudes encountered in his early measurements of turbulent fluctuations \cite{taylor1970some}).  Taylor launched his statistical theory of turbulence with the following motivation \cite{taylor1915eddy}:

\begin{quote}
The treatment of eddy motion in
either incompressible or compressible fluids by means of mathematics has always
been regarded as a problem of great difficulty, but this appears to be because
attention has chiefly been directed to the behaviour of eddies considered as individuals rather than to the average effect of a collection of eddies. The difference
between these two aspects of the question resembles the difference between the
consideration of the action of molecule on molecule in the dynamical theory of gases,
and the consideration of the average effect, on the properties of a gas, of the motion
of its molecules. 
\end{quote}

\noindent
Remarkably, at this outset of the statistical description of turbulence, Taylor then raised an issue that remains misunderstood to this day, and is a major topic of ongoing research.  Specifically in the same paper, Taylor discusses turbulent phenomena in the context of the diffusion of heat and momentum in gases.  In a note at the end of the paper, he is led to a discussion of the onset of turbulence, and the question of the stability of laminar flow.  At this time, Lord Rayleigh had developed some of the foundations for what we call today \lq\lq linear stability analysis\rq\rq, and Taylor \cite{taylor1915eddy} was exercised by the apparent conflict between Rayleigh's calculations of plane Poiseuille flow \cite{rayleigh1879stability,rayleigh1892viii} and Osborne Reynolds' experimental findings on pipe flow \cite{reynolds1883xxix}.  Specifically, Taylor was attempting to reconcile the fact that a laminar flow might be linearly stable in terms of its stability eigenvalues, as calculated by Rayleigh, even though in practice there is a transition to turbulence when the viscosity is lower than a critical amount, as observed by Reynolds.  Taylor assumed (correctly) that the laminar-turbulent transition has universal characteristics, so that it was appropriate to compare these different geometries, and attempted to argue for something like a nonlinear calculation of the response of a fluid to a perturbations.  He writes \cite{taylor1915eddy}:

\begin{quote}
In order that instability may be set up this
momentum must be absorbed by the walls.
There seems to be no particular reason
why an infinitesimal amount of viscosity should not cause a finite amount of momentum
to be absorbed by the walls.
\end{quote}

\noindent
Switching to a discussion of the momentum of gas molecules in the atmosphere being transferred to the ground, Taylor continues \cite{taylor1915eddy}:

\begin{quote}
... a
very large amount of momentum is communicated by means of eddies from the
atmosphere to the ground.
This momentum must ultimately pass from the eddies
to the ground by means of the almost infinitesimal viscosity of the air.
The actual
value of the viscosity of the air does not affect the rate at which momentum is
communicated to the ground, although it is the agent by means of which the
transference is effected.
In any case it is obvious that there is a finite difference, in regard to slipping at
the walls, between a perfectly inviscid fluid and one which has an infinitesimal
viscosity ... The finite loss of momentum at the walls due to an infinitesimal viscosity may be
compared with the finite loss of energy due to an infinitesimal viscosity at a surface
of discontinuity in a gas \cite{taylor1910conditions}
\end{quote}

With these mathematically imprecise, but prescient comments \cite{eyink2022onsager}, Taylor not only launched the statistical approach to fluid turbulence, but at the same time was the first to draw attention to a physical phenomenon what would turn out to be perhaps the most profound aspect of fully-developed turbulence: the dissipative anomaly.  In particular, Taylor was postulating, without mathematical proof, that the mere existence of an infinitesimal viscosity leads to finite and qualitatively different phenomenology than the case with zero viscosity.  Indeed, Taylor's assertions would later be formulated in mathematical language during the development of singular perturbation theory nearly 50 years later \cite{proudman1957expansions}, as a response to the challenge posed by Stokes' paradox \cite{stokes1851effect,veysey2007simple} for low Reynolds number flow around a sphere or cylinder.  

Furthermore, Taylor explicitly postulates that the value of the viscosity does not quantitatively influence the phenomenon that arises (momentum transfer); but, on the other hand, he asserts that the mere existence of non-zero viscosity is a necessary and sufficient condition for the phenomenon to occur.  

This claim is a very difficult one to prove.  
It echoes the phenomenon of \lq\lq scale interference\rq\rq\  that the renormalization group demonstrates is responsible for the existence of anomalous critical exponents in the field theoretic description of continuous phase transitions \cite{goldenfeld1992lectures}. Here, the particular problem is the fact that the behavior of thermodynamic and correlation functions near a critical point show singular behavior that is impossible if conventional assumptions are made.  For example, fluctuations of an order parameter become correlated over a distance, known as the correlation length, that diverges close to the critical point.  So one would naively expect that dimensional analysis could be used to explain the way in which observables such as heat capacity or magnetic susceptibility diverge with power-law exponents as a function of the temperature difference from the critical point $T-T_c$.  This argument unexpectedly fails: not only do critical exponents not have the values that mean field theory and dimensional analysis would suggest, but their values are also universal, transcending material-specific parameters or even the nature of the system itself: the liquid-gas critical point and the paramagnet-ferromagnet critical point behave identically!  In fact, it turns out that the lattice spacing, on the scale of Angstroms, also needs to be included as a length scale in the dimensional analysis, even though it may be many orders of magnitude smaller than the correlation length!  The renormalization group explains in detail how this works, and that in most cases, the critical exponents do not depend on the value of the lattice spacing (hence universality).  On the other hand, the mere existence of the lattice spacing as a parameter is a necessary and sufficient condition for the unusual scaling properties observed at continuous phase transitions.  

Today it is understood that there is a precise connection between the renormalization group account of phase transitions, the peculiar asymptotic arguments that Taylor invoked in his discussion of momentum transport and viscosity, and the singularities arising in low Reynolds number viscous flow \cite{goldenfeld1992lectures,chen1996renormalization,veysey2007simple}.  These connections will resurface when we discuss spontaneous stochasticity and the dissipative anomaly below.

By 1935, Taylor's use of the correlation function methods introduced in his earlier works allowed him to conclude that at high enough turbulent intensity, the dissipation rate was proportional to the velocity fluctuation kinetic energy $(u^\prime)^2$ divided by the time scale $L/u^\prime$ by which energy left the  eddies at large scale $L$ and entered what we now call a cascade to small scales.  This results in the famous formula that the dissipation rate per mass of fluid 

\begin{equation}
\varepsilon = C \frac{(u^\prime)^3}{L}
\label{eq:taylor}
\end{equation}
where the coefficient $C \sim O(1)$ for sufficiently large Reynolds number, which does not involve the viscosity explicitly.

Although it is clear that the eventual mechanism of dissipation must involve the viscosity, Taylor implied that the mere existence of a non-zero kinematic viscosity $\nu$ was all that was needed to establish a steady state in which the energy input at large scales was eventually dissipated at small scales, and that the  dissipation rate itself was independent of $\nu$.  It is possible that his motivation for assuming this was by analogy to the dissipation rate in a pressure-driven laminar pipe flow or plane Poiseuille flow, where the dissipation rate is usually written in terms of wall stress, and thus naturally depends on the viscosity.  However, it can also be written in terms of the pressure drop and the mean flow rate, and in this formulation, the dissipation rate is apparently independent of viscosity. The reason is that for a fixed pressure drop, the mean flow rate depends on the viscosity, and becomes smaller at high viscosity. So in this sense, the mean flow rate adjusts to an equilibrium value that is viscosity dependent.  For an analogous cancellation to occur in turbulent fluids, the velocity gradients in the fluid must attain an equilibrium where they scale in root mean square as $O(1/\sqrt{\nu})$. This scaling implies divergence  of the velocity gradients for \lq\lq small\rq\rq\ $\nu$ but, as we discuss at length below, the apparent singularity depends essentially on the reference length and time scales adopted. 

\section{Statistics of the laminar-turbulent transition}
\label{sec:trans}

Taylor's appeal to the existence of what one might call \cite{eyink2022onsager} a \lq\lq momentum anomaly\rq\rq\ was in the service of resolving the discrepancy between Rayleigh's and Reynolds understanding of the laminar-turbulent transition.  Today, we understand that the actual reason for the discrepancy is that the laminar-turbulent transition in the geometry being considered is sub-critical; with sufficient care, pipe flow can be made to stay laminar up to Reynolds numbers of order $10^5$ \cite{Pfenninger1961}.  Indeed, it is widely accepted that there is no linear instability of the laminar state to a turbulent one at any Reynolds number \cite{linearstability_pipe,meseguer2003linearized}.  In fact, even with laminar flows that do exhibit a linear instability at a finite Re, such as in Taylor-Couette flow with stationary inner cylinder, the linear instability occurs at a Re much larger than the observed sub-critical transition to turbulence, and so is irrelevant in practice.  

In these sub-critical transitions, it has become clear in the last 20 years that the turbulent behaviour is very different from what would be expected from conventional bifurcation theory approaches (for a recent survey of experimental results, see \cite{hof2023directed}).  In a sub-critical transition, the amplitude of the instability jumps discontinuously and hysteretically as the control parameter (ie. the Reynolds number) is varied. In such a transition there is a value of the control parameter (Reynolds number, $\Re_c$) below which the instability (to the turbulent state) does not occur. Slightly above this value, it is found experimentally that the turbulent fraction $\rho$ in the system reaches a steady state at long times, and the value of $\rho$ rises continuously from zero for $\Re > \Re_c$.  It is the continuous nature of $\rho (\Re)$ that permits the language and techniques \cite{goldenfeld1992lectures} of \lq\lq second-order phase transitions\rq\rq\ to be applied to the sub-critical laminar-turbulent transition!  For example, in the laminar-turbulent transition of quasi-one-dimensional circular Taylor-Couette flow \cite{hof_dp_1d_couette}, experiment yields $\rho \sim (\Re - \Re_c)^\beta$, where $\beta = 0.276$.  There is also strong evidence for scale invariance at the critical Re, data collapse of the time-dependent turbulent fraction $\rho (t, \Re)$, all consistent with scaling exponents characteristic of the directed percolation universality class.  

Why directed percolation?  Early numerical experiments on coupled map lattices had revealed spatiotemporal patterns of intermittency whose potential relevance to the coexistence of patches of laminar and turbulent regions in transitional pipe turbulence  was specifically noted by Kaneko \cite{kaneko1984period}.  Pomeau \cite{pomeau} noted that such patches of turbulence could spread and infect laminar regions of fluid, but the reverse process was inhibited by the linear stability of the laminar state, and suggested that this process might be in the universality class of directed percolation, as would be consistent with a general conjecture of Janssen and Grassberger \cite{janssen1981nonequilibrium,grassberger1982phase}. Chat{\'e} and Manneville \cite{chate1987transition} attempted to remove any possible artifacts due to the coupled map lattice description of spatiotemporal intermittency by studying a continuum partial differential equation --- the Kuramoto-Sivashinsky equation --- and noted analogies between the patterns of intermittency they observed and directed percolation.  For summaries of other early work, see \cite{hof_dp_1d_couette,ppmodel,barkley2016theoretical,manneville2016transition,goldenfeld2017statistical,goldenfeld2017turbulence,hof2023directed,avila2023transition}).

A major impetus to the field came from the seminal experiments of Hof and collaborators on the way in which turbulent puffs decay \cite{hof2008repeller} and split \cite{avila2011onset}. Back in 1883, Reynolds had first documented the presence of what he called \lq\lq flashes\rq\rq\ of turbulence, appearing interspersed with laminar regions of flow.  The Hof experiments established that below a critical Reynolds number $\Re_c\approx 2000 $ puffs decayed with a lifetime $\tau_d$ that grew as $\exp(\exp(\Re))$ up to constants of order unity, and above this value, puffs split with a lifetime that fell as $\exp(\exp(-\Re))$ up to constants of order unity.  The superexponential functional form of these findings could be understood from considerations of extreme value statistics \cite{goldenfeld2010extreme}, and were reported in other geometries too, such as channel flow \cite{shimizu2019exponential}.  Although we will not delve into it here, directed percolation has also been verified in 2D laminar-turbulent transitions, experimentally in Taylor-Couette flow \cite{klotz2022phase}, and computationally in a simplified model of channel flow known as Waleffe flow \cite{chantry2017universal}.

A mechanistic theory for these observations in pipe flow was developed by Shih et al. \cite{ppmodel}, whose direct numerical simulations of a single puff flow domain revealed that close to the critical Reynolds number, fluctuations in the Reynolds stress excited an azimuthal mean flow (termed a \lq\lq zonal flow\rq\rq\ because of its presence near the zero wavenumber part of the Fourier decomposition of the flow velocity) which in turn sheared and suppressed the turbulent fluctuations, giving rise to an activator-inhibitor or predator-prey dynamics. The Shih et al. theory recapitulated the superexponential form for the lifetime statistics observed in transitional pipe turbulence, without recourse to generic extreme value theory arguments. However, it also made another prediction.  The statistical theory of the energy flow in these predator-prey modes could be mapped into a stochastic field theory, and related to the directed percolation universality class using results already in the literature \cite{mobilia2007}.  This is the only direct prediction for the directed percolation transition in pipe flow that originates in the Navier-Stokes equations, and follows the standard process in statistical mechanics to make a prediction about the nature of phase transitions.   This prediction of directed percolation critical scaling was published alongside the observations of Hof and collaborators reporting their observations of directed percolation critical behaviour in the circular Taylor-Couette flow geometry \cite{hof_dp_1d_couette}.

As usual with a phase transition \cite{goldenfeld1992lectures}, the predator-prey modes observed by Shih et al. are necessarily weak, since they occur in the vicinity of a critical point; but they have a non-perturbative impact on the singular behavior.  A similar dynamics had previously been predicted using heuristic physical arguments by Diamond and collaborators \cite{diamond1994} in the context of the Low-High Confinement transition in tokomaks, and after much controversy the predator-prey modes were finally observed nearly 20 years later \cite{estradaPRL2011}.  For transitional turbulence, a heuristic mean field theory derivation of the predator-prey modes found in \cite{ppmodel} was provided in \cite{goldenfeld2017turbulence}.  In transitional pipe turbulence, the predator-prey modes exhibited by the Navier-Stokes solutions, were later shown to couple to the streamwise flow \cite{chen2024mean,khan2024examination} and to strongly influence the transitional behavior through the self-sustaining process \cite{waleffe1997self}, widely accepted to play a major role in the flow dynamics near the transition.  Later, the predator-prey theory for the energetics of transitional flow in quasi-one-dimensional geometries was extended to include the background mean flow that is the source of the turbulent kinetic energy \cite{wang2022stochastic}, predicting not only the behaviour in circular Taylor-Couette flow but also the detailed dynamics of regions of expanding turbulence known as \lq\lq slugs\rq\rq\ that are observed at Reynolds numbers above the puff transitional region \cite{Wang2025stoch}.  

Finally, the reader might wonder why it is that in pipe flow, superexponential scaling is observed, whereas power-law critical scaling behaviour is found in circular Taylor-Couette flow.  It turns out that the extended stochastic field theory for the predator-prey dynamics can be analyzed in terms of the Ginzburg criterion for phase transitions \cite{ginzburg1958light,levanyuk1959contribution,ginzburg1961some,goldenfeld1992lectures}, with the result that the critical region is much smaller in the pipe geometry than the circular Taylor-Couette geometry, due to the difference between energy injection in the streamwise direction along the pipe and energy injection in the radial direction for the Taylor-Couette geometry \cite{Wang2025SizeCriticalRegion}.  For the pipe flow case, the streamwise localization of puffs acts to amplify the finite-size effect on their lifetime fluctuations, which leads to the superexponential behavior \cite{Shih2026}. 

This narrative has focused on the way in which the intrinsic statistical fluctuations were observed experimentally and predicted theoretically near the laminar-turbulent transition.  But we have left out one aspect that has not received sufficient attention: the existence of an anomaly in the mathematical description of the phase transition.  We would like to end this section with some remarks about unresolved theoretical issues related to stochasticity and anomalies.  In quantum field theory, an anomaly is said to occur when a symmetry of a classical Lagrangian is not preserved under quantization \cite{bertlmann2000anomalies,fujikawa2004path}.  In the theory of the laminar-turbulent transition, the mechanistic explanation for directed percolation stems from the fact that the stochastic field theory of the important modes near the transition is of the predator-prey or activator-inhibitor type. A simple description of a predator-prey model for the internal dynamics of a single puff, neglecting space, uses the Lotka-Volterra equations:
\begin{align}
    &\frac{dA}{dt}=pAB-dA\\
    &\frac{dB}{dt}= bB(1-B/\kappa) -pAB
\label{eq:pp}
\end{align}
where $A$ and $B$ are density of predator and prey respectively, $p$ is predation rate, $d$ is death rate of predators, $b$ is the birth rate of prey, and $\kappa$ is the so-called carrying capacity of the ecosystem.  When $B=\kappa$, the growth rate of the prey density becomes zero, meaning that the prey have utilised all the resources of the ecosystem.  In the case of the transition to pipe turbulence, the ecosystem is the energy provided by the incoming mean flow, the predator $A$ is the energy of the zonal or azimuthal flow, the prey $B$ is the energy of the small-scale turbulence \cite{goldenfeld2017turbulence}.  These classical equations are supposed to capture the population cycles wherein prey population grows due to availability of resources (energy from the background mean flow), thus providing food for the predators whose population then grows.  This causes the prey population to decline, and thus eventually the decline of the predator population too due to starvation.  Once the predator population has declined, the prey population is free to grow and the cycle begins again.  Such oscillations might sound like a limit cycle, with a $\pi/2$ phase delay between prey and predator.  But in fact the Lotka-Volterra equations Eq.~(\ref{eq:pp}) do not have cyclical solutions at all!  They decay to a fixed point not a limit cycle, when $\kappa < \infty$.  For the case $\kappa = \infty$, the equations have cyclic solutions but they are not structurally stable with respect to perturbation (e.g. $\kappa \rightarrow \infty$ and $\kappa = \infty$ yield qualitatively different solutions: cyclic vs. constant).  Moreover in this case, the cyclic solutions have a conserved quantity and their behaviour depends on the initial conditions, so there is no limit cycle attractor at long times.  In summary, these deterministic equations have no physically acceptable solutions that account for population cycles!

One might fix this problem by postulating additional levels of realism, such as that when the prey population is sufficiently plentiful, the dynamics of predator and prey meeting should be independent of prey density, rather than the law of mass action expression that the rate is proportional to $A \times B$.  One way to do this is to replace $p \rightarrow p/(C+B)$ in the equation, where $C$ is a constant.  Then, for large $B \gg C$, the predation rate only depends on the density of predators.  This does indeed allow the equations to have a limit cycle solution.  

However, this is not the solution of the problem, as can be seen by replacing the continuum Lotka-Volterra equations by equations for the {\it number} of predators and prey rather than their densities.  The numbers $N_A$ and $N_B$ are integer valued, and the corresponding equations are of the form:
\begin{align}
    &N_A\xrightarrow{d} \phi\\
    &N_A+N_B\xrightarrow{p} 2N_A\\
    &N_B\xrightarrow{b} 2N_B\\
    &2N_B\xrightarrow{b/\kappa}N_B
\label{eq:stochpp}
\end{align}
where the rate constants for the individual reactions are given above the arrows.  Simulation of these discrete equations yields persistent population cycles!  But the cycles are not deterministic.  Instead they have stochastic fluctuations with a power spectrum that can be calculated using the van Kampen expansion, including when spatial degrees of freedom are also included \cite{mckane2005predator,butler2009robust,tauber2012population}. The take home message from this brief description is that exactly the same physics, when expressed in a deterministic formalism or a discrete stochastic formalism, leads to completely different phenomenological outcomes.  The stochastic process described by Eq.~(\ref{eq:stochpp}) has as its mean field equation the deterministic Lotka-Volterra equations Eq.~(\ref{eq:pp}) but it is necessary to include the one-loop corrections of $O(1/\sqrt{N})$ (where $N$ is either $N_A$ or $N_B$) to the equations to capture the population cycles.  In this sense, there is no meaningful deterministic limit of the stochastic Lotka-Volterra equations, when they are quantized in the sense that the variables are integer-valued rather than real-valued.  

This situation mirrors what happens in certain quantum field theories, not just the stochastic field theories that we are considering here.  In quantum field theory, the issue arose in considering the phenomenon of the decay into two photons of a particle known as the neutral pion. The rate of decay can be calculated from quantum electrodynamics, and when first attempted \cite{steinberger1949use,fukuda1949gamma} yielded a non-zero result that was consistent with the available experiments.  However, the calculation was not fully gauge-invariant and so it was clear that there was something missing.  Moreover, it was known that the appropriate theoretical model, (e.g.) massless chiral fermions in $3+1$ dimensions coupled to the electromagnetic field, has a classical Lagrangian that conserves a Noether symmetry known as chiral symmetry (essentially that there is a symmetry between left and right-handed vectors in the theory). This classical symmetry would forbid the pion decay.  Thus, how does the decay happen in Nature?  And how to calculate its rate correctly without breaking gauge invariance?  It is now understood that this phenomenon arises because the correct process of quantizing the theory and maintaining gauge invariance can only be done if one also breaks the chiral symmetry \cite{adler1969axial,bell1969pcac}.  In other words, there is no meaningful classical deterministic limit of the quantum equations for chiral fermions.  This phenomenon --- the chiral anomaly --- was a huge shock when it was discovered in high energy physics.  Many other anomalies have been identified since then \cite{bertlmann2000anomalies,fujikawa2004path}, and this is an active field today in condensed matter and high energy physics \cite{arouca2022quantum,mcgreevy2023generalized}. 

The stochastic anomaly in the description of the laminar-turbulent transition is discussed in more detail elsewhere \cite{Wang2026anomalies}, but is only one of the several anomalies that arise in turbulence, as we now discuss.  

\section{Turbulence anomalies}
\label{sec:diss}

\subsection{Dissipative anomaly}
\label{subsec:anom_HIT}
The first anomaly discovered was not in fact the chiral anomaly in quantum field theory.  It was the dissipative anomaly in fluid turbulence, and can be traced to the famous 1949 paper by Onsager on statistical hydrodynamics \cite{onsager1949statistical}.  
Since one of us (GE) recently published a very comprehensive review of Onsager's theory and its current 
status \cite{eyink2024onsager}, we shall here be more brief and add just a few comments necessary
to put Onsager's work into the context of the present essay. 

The roots of the dissipative anomaly lie in Eq.~(\ref{eq:taylor}) from G.I. Taylor, describing how the dissipation rate in fully-developed turbulence is independent of viscosity $\nu$. Onsager cited Taylor's papers from the 1930's and especially the 1943 experimental study of Dryden on grid-turbulence in a wind-tunnel \cite{dryden1943review}, which attempted to evaluate the constant prefactor $A$ in Taylor's relation $\varepsilon \sim A u^{\prime\, 3}/L.$ In fact, it was not until the study of K. R. Sreenivasan 
forty years later \cite{sreenivasan1984scaling} that convincing evidence was compiled that 
$A$ becomes Reynolds independent in grid turbulence at $\Re\gg 1.$ Of course, other data in wake 
flows has long been available with similar implications, e.g. from the drag coefficients 
$C_D(\Re):=F_D/[(1/2)\rho U^2 A]$ of solid bodies, defined by non-dimensionalizing the drag force $F_D$
with the fluid mass density $\rho,$ velocity $U$ of the body, and $A$ its cross-sectional area. 
It is a standard observation that drag coefficients of bluff bodies have positive constant
values at high Reynolds number and this $\Re$-independence for $\Re\gg 1$ implies Taylor's scaling of 
energy dissipation in the wake. In fact, $F_D U$ is 
the power expended to move the body, which is ultimately dissipated in the turbulent wake with characteristic
volume $\sim A^{3/2}$ and integral scale $L\sim A^{1/2}$, so that 
$\varepsilon/(U^3/L) \sim \frac{F_D U/(\rho  A^{3/2})}{U^3/A^{1/2}} \sim C_D.$
See \cite{frisch1995turbulence}, \S 5.2.
A very clean example 
of such $\Re$-independence is the drag coefficient of a circular disk with face normal to the flow, 
which Prandtl's group  \cite{wieselsberger1923ergebnisse}, p.29, and independently Shoemaker \cite{shoemaker1926resistance} had measured in the 1920's up to $\Re\simeq 5\times 10^6$
and found essentially constant for four decades of Reynolds number. 

However, the experimental picture in Onsager's day and up to the present time is much richer 
and more complex than the above observations alone suggest. Onsager in his 1949 work cited a paper 
on turbulent pipe flow by R. B. Montgomery \cite{montgomery1943generalization} which he very 
enthusiastically recommended to colleagues (see \cite{eyink2006onsager}). The main point of 
Montgomery's paper was to develop a new theory of mixing length and the Darcy friction factor 
$f(\Re):=D (\Delta P/l)/[\rho U^2/2]$, with $D$ the pipe diameter, $(\Delta P/l)$ the downstream
pressure gradient, and $U$ the fluid bulk velocity, 
which Montgomery compared favorably with experimental data 
of Nikuradse \cite{nikuradse1932laws,nikuradse1933laws}. Note that the friction factor 
measures also the energy dissipation rate in a pipe section of length $l,$ since 
$\Delta P A U$ is the energy input by pressure work and thus $\varepsilon/(U^3/D)\sim 
\frac{\Delta P A U/(\rho A l)}{U^3/D}\sim f.$ What is most interesting about Nikuradse's 
results is that he found $f(\Re)\sim \Re^{-1/4}$ as $\Re\to \infty$ for pipes with hydraulically 
smooth walls  \cite{nikuradse1932laws} but instead $f(\Re)\to f_\infty>0$ as $\Re\to \infty$ for pipes with 
hydraulically rough walls and the constant $f_\infty$ was in that case 
increasing with the roughness height \cite{nikuradse1933laws}. 
Thus, the scaling of energy dissipation suggested by Taylor only leads to a $\nu$-independent result in the rough-wall case, but not in the smooth-wall case! This is not true 
in the case of wake flows past bluff bodies where, even with highly polished surfaces, the 
drag coefficient $C_D$ appears to become a non-zero constant at $\Re.$ This dichotomous 
situation persists to the present day. In several other internal flows, such as Taylor-Couette 
or von K\'arm\'an flows, it is found that $\varepsilon/(U^3/L)\to 0$ as $\Re\to\infty$ with 
smooth walls, but that $\varepsilon/(U^3/L)\to D_*>0$ with rough walls and $D_*$ increases 
with roughness \cite{cadot1997energy}. On the other hand, in external wake flows past 
cylinders, plates, grids, etc. it appears that  $\varepsilon/(U^3/L)\to D_*>0$ as $\Re\to\infty,$ 
even with hydraulically smooth walls \cite{pearson2002measurements}.

Onsager, as we have noted, was certainly aware of this somewhat complex experimental situation 
but he did not attempt to explain when Taylor's scaling law for energy dissipation rate is 
valid or not. Instead, Onsager deduced important consequences of Taylor's scaling 
{\it whenever it holds} and in particular for wake flows. The starting point of Onsager's 
analysis was the observation 
that the local viscous dissipation rate per mass is $\varepsilon(\bx,t)=\nu|\grad\bu|^2.$ Thus,
$\varepsilon(\bx,t)/(U^3/L)=(1/\Re)|\hat{\grad}\hat{\bu}|^2$ where quantities with hats 
have been non-dimensionalized with outer scales $L$ and $U$. It therefore 
becomes obvious that $\varepsilon/(U^3/L)$ can have a finite positive limit as 
$\Re\to \infty$ only if $|\hat{\grad}\hat{\bu}|\to\infty$ in that limit. 
In Onsager's own words, ``the dissipation of energy is regarded as primarily a 
`violet catastrophe'.'' \cite{onsager1945distribution}. In more modern language, 
there are {\it ultraviolet divergences} in the limit $\Re\to\infty$ so that 
dimensionless velocity-gradients can no longer exist pointwise as ordinary functions 
but only as generalized functions or distributions. This is the same situation 
as in quantum field theory, where continuum quantum fields are not defined pointwise but 
only when smeared with test functions, i.e. they are operator-valued distributions. 

It is interesting, by the way, that this conclusion of Onsager's holds even 
when $\varepsilon/(U^3/L)\sim \Re^{-p}$ as $\Re\to\infty$ with $0<p<1.$ To use a 
felicitous terminology of Bedrossian et al. \cite{bedrossian2019sufficient}, the 
situation with $p=0$ (Taylor's scaling) is a {\it strong anomaly} whereas $0<p<1$
corresponds to a {\it weak anomaly}, for which dimensionless dissipation rate 
vanishes as $\Re$ but slower than the laminar rate $\sim \Re^{-1}.$ Experimentally,
all of the turbulent flows discussed above exhibit at least a weak dissipation anomaly. 
Another important remark necessary to present confusion is that the ``divergence''
of the dimensionless velocity-gradients does not necessarily threaten the validity 
of a macroscopic hydrodynamic description, because velocity-gradients in inner 
scale units can remain finite! Thus, if one defines the Kolmogorov dissipation 
length $\eta=\nu^{3/4}/\varepsilon^{1/4}$ and Kolmogorov velocity $u_\eta=(\nu\varepsilon)^{1/4}$
in terms of the mean dissipation $\varepsilon,$ then $\varepsilon(\bx,t)/\varepsilon
=|\tilde{\grad}\tilde{\bu}|^2$ and tilde means that the quantities have been 
non-dimensionalized with $\eta$ and $u_\eta.$ Thus, the gradients in these units 
can only diverge if there are large fluctuations in the local dissipation rate 
as $\Re\to \infty.$ This is the intermittency phenomenon that we will discuss later 
but, barring extreme intermittency, these gradients remain finite as $\Re\to\infty$. 
It is crucial to keep in mind that $\Re=UL/\nu$ is made unboundedly large not by increasing 
$U$ (which would violate incompressibilty) or by decreasing $\nu$ (which is a material constant),
but instead by increasing $L$: think of blue whales and jumbo jets! It is only the gradients 
measured on these large scales which must diverge.  
 
It is easy to non-dimensionalize the Navier-Stokes equations in outer units,
which then have the form 
\be \partial_{\hat{t}}\hat{\bu}+(\hat{\bu}\bdot\hat{\grad})\hat{\bu}
=-\hat{\grad}\hat{p}+(1/\Re)\hat{\triangle}\hat{\bu}, \quad 
\hat{\grad}\bdot\hat{\bu}=0. \lb{NS-nondim} \ee 
For convenience, we always assume hereafter non-dimensionalization in 
outer units and drop the hats. 
Taking $\Re\to\infty,$ one would guess that the limiting velocity field
satisfies the incompressible Euler equations. However, Onsager's first 
observation implies that this limit statement cannot be true in the na\"ive sense,
because the gradients involved would diverge. What Onsager shrewdly realized 
is that the Euler equations could hold for the limit ``in a generalized description''
\cite{onsager1949statistical} and that Euler solutions in this generalized 
sense no longer need conserve kinetic energy! In fact, Onsager's suggested 
generalization corresponds exactly to the modern notion of {\it weak solution}
so that the limiting dimensionless velocity would satisfy 
\be \partial_t\bu+\grad\bdot(\bu\bu) 
=-\grad p, \quad  \grad\bdot \bu =0. \lb{Euler-weak} \ee 
in the sense of distributions \cite{eyink1994energy,delellis2013continuous}. 
To derive a balance equation for kinetic energy in this framework 
is delicate, because standard manipulations, e.g. with partial-differentiation,
are ill-defined. Just as in quantum field-theory, the quantities must be 
regularized to eliminate UV divergences to make calculations well-defined. 
Although Onsager never published his derivations and only disseminated 
his results in 1945 via private letters \cite{eyink2006onsager}, it is now known 
that he used a {\it point-splitting regularization} by introducing a 
displacement vector $\br$ to give a kinetic energy density 
$(1/2)\bu(\bx+\br,t)\bdot \bu(\bx,t)$
with UV divergences removed. In this way, by taking time derivatives first 
and then subsequently the limit $r\to 0,$ Onsager derived a distributional 
kinetic energy balance 
\be \partial_t\left(\frac{1}{2}|\bu|^2\right)
+\grad\bdot\left[\left(\frac{1}{2}|\bu|^2+p\right)\bu\right] 
=-D(\bu). \lb{Euler-anom} \ee 
where the {\it dissipative anomaly} term has the limit expression 
\be D(\bu) = - \lim_{r\to 0} \frac{3}{4r} \langle (\hat{\br}\bdot\delta\bu(\br)) 
|\delta\bu(\br)|^2\rangle_{{\rm ang}} \lb{D-anom} \ee 
where $\delta\bu(\br;\bx,t)=\bu(\bx+\br,t)-\bu(\bx,t)$ and the bracket 
$\langle\cdot\rangle_{{\rm ang}}$ denotes spherical angle average over the 
direction vector $\hat{\br}=\br/r.$ In fact, Onsager in his unpublished 
notes derived this result only in space-integrated form and the result as presented 
above was independently rediscovered and extended to space-time local 
form by mathematicians Duchon \& Robert 55 years later \cite{duchon2000inertial}. 
It is historically interesting that 
the first derivation of the chiral anomaly in QED by Julian Schwinger in 1951 used a 
very similar point-splitting regularization \cite{schwinger1951gauge}, although 
Schwinger did not seem to fully appreciate that his result violated conservation 
of chiral charge \cite{adler2005anomalies}. Onsager clearly understood the implication 
that kinetic energy is not conserved, but his published and private remarks 
were also not appreciated until decades later. 

An immediate consequence of \eqref{D-anom}, as pointed out by Onsager in the final
paragraph of his 1949 paper \cite{onsager1949statistical}, is that a strong dissipative 
anomaly with $D(\bu)>0$ requires that the velocity field $\bu$ obtained in the 
limit $\Re\to\infty$ must develop H\"older singularities with exponent $h\leq 1/3.$
Indeed if one assumes that 
\be |\delta\bu(\br)|\leq C |\br|^h \lb{Hoelder} \ee
for any $h>1/3,$ then $\hat{\br}\bdot\delta\bu(\br)
|\delta\bu(\br)|^2/r=O(r^{3h-1})$ and necessarily $D(\bu)\equiv 0.$ Remarkably, 
Onsager made a precise prediction about turbulent singularities in the inertial 
range of scales deduced from the empirical validity of Taylor's relation. Thus, 
Onsager's ideas are a clear forerunner of the Parisi-Frisch multifractal model \cite{frisch1985singularity} which postulates
an entire multifractal spectrum $D(h)$ of such H\"older singularities. In that 
framework, Onsager's prediction corresponds to the statement that $h_{\min}\leq 1/3$
and has been amply confirmed by subsequent experiments. For example, see 
the careful study of Lashermes et al. \cite{lashermes2008comprehensive}, which analyze 
grid turbulence data up to Taylor Reynolds numbers $\Re_\lambda\simeq 2500$ from 
the Modane wind-tunnel using a wavelet method. 

To head off possible misunderstanding, it is important to emphasize that the 
turbulent dissipative anomaly does not in any way violate conservation of energy. 
The latter arises from fundamental principles, since the 
microscopic particle dynamics are Hamiltonian and energy conservation arises 
by Noether's theorem from their time-translation invariance. However, there is no 
fundamental physical principle of ``conservation of kinetic energy''! In fact, 
it was proved by Duchon \& Robert \cite{duchon2000inertial} that the dissipative 
anomaly can also be obtained directly from the kinetic energy balance for 
Navier-Stokes, with 
\be D(\bu) = \lim_{\nu\to 0} \nu|\grad\bu|^2. \lb{visc-anom} \ee 
Thus, the expression \eqref{D-anom} representing kinetic energy cascade through 
the inertial range matches on exactly to the viscous dissipation of kinetic energy 
\eqref{visc-anom}, which is converted into heat or internal energy of the constituent 
molecules of the fluid. Thus, total energy is conserved. The reason that the 
kinetic energy balance \eqref{Euler-anom} is considered ``anomalous'' is that the
incompressible Euler equations are (non-canonical) Hamiltonian PDE's and the 
kinetic energy is the Hamiltonian \cite{salmon1988hamiltonian,morrison2006hamiltonian}. 
One would na\"ively expect a Hamiltonian dynamics to conserve its own Hamiltonian, 
but that is not necessarily true for singular, infinite-dimensional systems like 
Euler equations! The same remark is generally true for conservation laws of the 
ideal fluid equations, such as helicity conservation or Kelvin circulation theorem,
that arise by Noether's theorem from symmetries of Hamiltonian fluid dynamics 
\cite{salmon1988hamiltonian,morrison2006hamiltonian}. These are {\it emergent 
hydrodynamic conservation laws} which do not have any microscopic analogues 
and they can be potentially vitiated by anomalies in singular limits such as $\Re\to\infty.$
To make clear the close connection with quantum anomalies, the limit $\Re\rightarrow \infty$
is the analogue of the semi-classical limit $\hbar\to 0$ of a quantum field theory, 
whilst setting $\Re=\infty$ is the analogue of simply setting $\hbar=0$ 
in quantum field theory, i.e. the na\"ive classical limit.


In addition to his $1/3$-H\"older prediction, Onsager also suggested that the observed dissipation 
in turbulent flows might be idealized in the regime $\Re\gg 1$ by dissipative Euler solutions, 
writing that ``turbulent dissipation as described could take place just as readily without the final assistance by viscosity'' \cite{onsager1949statistical}. It is now rigorously proved that 
dissipative, H\"older-continuous Euler solutions as conjectured by Onsager do exist, with 
exponents right up to the critical value $h=1/3.$ This mathematical development came from 
a surprising application of \lq\lq convex integration\rq\rq\  methods arising from work of John Nash, 
which was lead by C. De Lellis and L. Sz\'ekelyhidi, Jr. \cite{delellis2009euler,delellis2010admissibility}
and culminated in the construction up to the critical exponent 
\cite{isett2018proof,buckmaster2019onsager}. To learn more about this important 
development, see our recent review \cite{eyink2024onsager} or especially the 
reviews by the mathematicians directly involved \cite{delellis2013continuous,delellis2019turbulence}.
Here we just emphasize one very surprising conclusion of this work, the profligate 
non-uniqueness of dissipative, H\"older-continuous weak Euler solutions even for given 
fixed initial data. See \cite{delellis2010admissibility} and 
\cite{daneri2021non,delellis2023nonuniqueness,isett2022nonuniqueness,giri2023L3}
for recent developments. To quote just one such result \cite{daneri2021non},
there exist initial data $\bu_0$ which are H\"older continuous with exponent $h=1/3-\epsilon$ and dense 
in the set of divergence-free, finite-energy fields, so that for each of these 
initial data there are infinitely-many weak Euler solutions with decreasing kinetic 
energy and H\"older continuous with exponent $h'=h-\epsilon$. This type of Nash ``non-rigidity'' 
or ``flexibility'' of weak Euler solutions is very different from solutions of scalar 
hyperbolic conservation laws, like Burgers equation, which are unique subject 
to a dissipation condition. We shall mention possible physical implications further below. 

There are many questions still left completely open in this area, which bring 
us to the frontier of current research. Can dissipative Euler solutions be 
obtained in the infinite Reynolds-number limit of Navier-Stokes? Are dissipative 
anomalies strong or weak? What is the origin of the turbulent singularities? 
How does one explain the empirical observations on the role of solid walls? 
We will offer just a few remarks and personal views. We are convinced that 
interactions of turbulence with solid walls are crucial to produce a strong 
dissipative anomaly, at least for low Mach-number incompressible flow. As one 
piece of evidence, we cite the recent numerical study of forced, homogeneous, isotropic 
turbulence in a periodic box whose results are in better agreement with a weak 
anomaly \cite{iyer2025turbulence}. In fact, to our knowledge, all flows that give 
substantial evidence of a strong dissipative anomaly have the common feature
that drag and dissipation arise at high-$\Re$ due to ``form drag'' from pressure 
forces, both in wake flows past past bluff bodies such as spheres 
\cite{achenbach1972experiments} and internal flows through pipes or channels
with rough walls \cite{busse2017reynolds}. Wall-bounded flows that exhibit 
only weak anomalies \cite{nikuradse1932laws,cadot1997energy} have not only
smooth walls but more specifically get no drag contribution from pressure forces. 
The classical Josephson-Anderson relation implies that drag must be associated 
with vorticity flux into the flow interior from the solid surface \cite{eyink2021josephson,kumar2024josephson}, and this flux must persist even in 
the infinite-Reynolds limit in order to produce a strong anomaly \cite{quan2024onsager}. 
The likely source of this vorticity flux is separation of thin vortex sheets 
from the walls, recalling the ideas of Taylor that high-Reynolds fluids form 
something like ``shock singularities'' at solid walls. Taylor's idea has 
recently gotten some support from Onsager's approach, as initiated by 
mathematicians Bardos \& Titi \cite{bardos2018onsager}. Realizing that 
diverging velocity-gradients at the wall are an additional source of UV 
divergences, these authors introduced a novel regularization that both filters 
out small-scales eddies in the interior and also windows out eddies near 
the wall. Using this approach, one of us (GE) has shown that velocity-discontinuities at the wall 
are necessary to produce Taylor's ``momentum anomaly'' \cite{quan2025inertial}.
Under the same discontinuity condition one can show that turbulent singularities may arise 
entirely at the wall in the infinite-$\Re$ limit, without the need of any blow-up of the smooth 
Euler solution \cite{eyink2025weak}. The above picture provides many experimentally testable predictions. 

\subsection{Anomalous dimensions}

In the introduction we highlighted the phenomenon of \lq\lq scale interference\rq\rq\ 
that is responsible for the existence of anomalous critical exponents in continuous phase transitions \cite{goldenfeld1992lectures}. In that case, the lattice spacing $\ell$, despite being 
orders of magnitude smaller than the correlation length, must be included as a length scale in the dimensional analysis, and enters into quantities such as the correlation function $G(k)$ expressed as a function of wavenumber $k$ in a singular way: $G(k) \sim \ell^{\phi}k^{-2+\phi}$, allowing the anomalous behaviour as a function of $k$: $G(k)\sim k^{-2+\phi}$, which  would seem to violate dimensional analysis, unless the dependence on $\ell$ is included. The exponent $\phi$ is an example of a so-called anomalous dimension. This necessity of including a very small quantity for correct 
dimensional analysis has been termed {\it self-similarity of the second kind}
\cite{barenblatt1972self,barenblatt1996scaling,goldenfeld1992lectures}, in contrast to the case of {\it self-similarity of the first kind}, where a small variable can be taken to zero without there being any singular behaviour.  And as Taylor presciently commented, this is an example where the existence of a small quantity, here the lattice spacing in the formula for $G(k)$, allows for the observed singular variation with respect to another variable, here $k$.  The anomalous dimension $\phi$ does not, however, depend on $\ell$.  But without the singular dependence on $\ell$, it would be impossible to have scaling exponents that were different from the ones predicted by mean field theory and dimensional analysis.  

In fact, a strong dissipative anomaly with Taylor's scaling $\varepsilon\sim (u^\prime)^3/L$ implies first kind similarity with respect to viscosity,
because the limit $\nu\to 0$ safely exists. This mirrors the situation in quantum field theory 
where the chiral anomaly is not itself renormalized by quantum corrections and thus cannot acquire 
an anomalous scaling dimension \cite{adler1969absence,zee1972axial}. However, in certain situations 
such as pipe flow the dissipation does exhibit self-similarity of the second kind, but with respect to 
wall roughness rather than viscosity. To explain this, we must dive deeper into the classical 
experimental results on pipe flow. 

During the years 1932-1933, J. Nikuradse published two seminal papers in which he measured 
Darcy friction factor $f$ as a function of $\Re$.  In the first paper \cite{nikuradse1932laws}, the pipe walls were nominally smooth.  In the second paper \cite{nikuradse1933laws}, the roughness of the pipe walls was varied by gluing monodisperse sand grains of radius $r$ to the walls, thus varying the relative roughness $r/D$ where $D$ is the pipe diameter. The second paper is the more important, in our opinion, particularly as this experiment has never been repeated, at least in three dimensional turbulence.  
For a laminar flow, 
$f=64/\Re$.  The laminar regime ends with the laminar-turbulent transition around $\Re=O(2\times 10^3)$, which leads to an increase in $f$, the so-called drag catastrophe.  For $ 3.6 < \log_{10} \Re < 4.8$, in the case of the smoothest pipe examined ($r/D=15$), Nikuradse found that $f \propto \Re^{-1/4}$.  This interval and the scaling law is attributed to Blasius \cite{BLAS13}.  For $\log_{10}\Re > 5.8$ the friction factor is constant with increasing Re.  This ultimate asymptotic regime is associated with Strickler \cite{STRI23}.  As the roughness ratio is increased, the upper limit of the Blasius regime decreases and the Strickler regime begins at a lower Re.  Overall in the Strickler regime, the friction factor $f \propto (r/D)^{1/3}$ and independent of Re.  In between the Blasius and Strickler regimes, the friction factor is non-monotonic, resembling a spoon shape. 

Nikuradse's data clearly shows that there is a dissipative anomaly at large $\Re$ but the magnitude of the anomaly is associated with the wall roughness.  Furthermore, as the roughness decreases, the extent of the Blasius regime increases.  One of us (NG) \cite{goldenfeld2006roughness} showed that if fully-developed turbulence is regarded as a non-equilibrium critical phenomenon at $\Re\rightarrow \infty$ a scaling law for the friction factor could be developed (termed \lq\lq roughness-induced criticality\rq\rq), and Nikuradse's data as a function of roughness ratio $r/D$ and Re could be collapsed onto a universal curve.  This argument implied that extrapolating to $r/D\rightarrow 0$ leads to the conclusion that the friction factor in an asymptotically smooth pipe will scale with the Blasius exponent.  
This is the scaling referred earlier as a weak anomaly, in the sense that $f$ vanishes  with $\Re$
but more slowly than the laminar result.   
For any small but non-zero roughness ratio $r/D$ the ultimate asymptotic regime of the friction factor will be the Strickler scaling, and the dissipative anomaly corresponding to a friction factor independent or Re.  Later Mehrafarin and Pourtolami \cite{mehrafarin2008intermittency} extended this analysis and showed that the data collapse could be improved if the intermittency correction to the turbulent velocity fluctuation energy spectrum $E(k)$ was included.  This remarkable result is a direct manifestation of a non-equilibrium fluctuation-dissipation relation, connecting the velocity fluctuations (through their intermittency exponent) with the dissipation (via the friction factor).  A heuristic derivation of just such a relation was given by Gioia and Chakraborty \cite{gioia2006turbulent}, showing explicitly how to compute the friction factor in terms of the energy spectrum; despite an oversimplified treatment of the velocity profile away from the boundary, this ingenious result correctly provides at mean field level the scaling laws for both Blasius and Strickler regimes, and satisfies the roughness-induced critical data collapse.

The Nikuradse experiments occupy a unique place in the literature of turbulence, due to their extensive coverage of Reynolds number and roughness, and the care with which they were performed.  As such they have never been repeated. However, a two-dimensional (2D) version of the experiments has been performed in a set of measurements that began to test the Gioia-Chakraborty predictions.  The point about 2D turbulent flows is that there are two cascades, one for energy and one for enstrophy, with different functional forms for the energy spectrum $E(k)$.  Thus, by creating these flows in turbulent soap films, it is possible to check the predictions \cite{guttenberg2009friction} for the 2D analogues of Blasius and Strickler regimes for both cascades.  These experiments clearly indicate that the enstrophy cascade has exponents in complete agreement with expectations, in both Blasius \cite{tran2010macroscopic,kellay2012testing} and Strickler \cite{vilquin2021asymptotic} regimes.  

Although energy dissipation has first kind similarity with respect to viscosity, 
this is not the case for all turbulence quantities. In particular, the intermittency
phenomenon mentioned above implies that higher powers of velocity-gradients, schematically 
$(\grad\bu)^{\otimes p}$ for $p>2$ depend upon Reynolds number, even when non-dimensionalized 
by the $p/2$ power of $\langle|\grad\bu|^2\rangle.$ E.g. see \cite{wyngaard1972some}. In fact, 
this means that such quantities have formally both a UV divergence as $\nu\to 0$ and 
and an IR divergence as $L\to\infty.$ Thirty years ago \cite{eyink1994analogies}, 
we argued that the limit $\nu\to 0$ and the limit $L\to \infty$ both suggest analogies 
with the continuum limit of quantum field theory, which are different but both correct and useful. 
In fact, in the intervening years the problem of anomalous scaling has seen major 
progress \cite{falkovich2001particles} in a model of passive scalar advection, the Kraichnan model \cite{kraichnan1968small}. As we predicted, both analogies can be used there and yield 
the same results \cite{kupiainen2007scaling}. Intermittency and anomalous scaling for 
Navier-Stokes remain hard problems. An operator-product expansion proposed for 
renormalized powers of velocity-gradients predicts multifractal scaling of velocity 
increments and relations between anomalous scaling exponents of those renormalized 
operators and the anomalous scaling exponents of velocity structure functions
\cite{eyink1993lagrangian}. This relation has been controversial but just recently 
new evidence in favor has emerged from well-resolved, high-$\Re$ numerical 
simulations \cite{buaria2025intermittency}. We hope to see further progress,
perhaps with creative new ideas. One very interesting recent discovery is IR and UV 
divergences in kinetic wave turbulence at higher-order in the nonlinearity 
\cite{rosenhaus2024wave,rosenhaus2024interaction,rosenhaus2025weak}, which 
may signal new anomalies in such systems. 

\section{Spontaneous stochasticity}
\label{sec:spontstoch}

In addition to the anomaly at transition discussed in Section \ref{sec:trans},
there is another phenomenon that has as well been aptly called a \lq\lq stochastic anomaly'' \cite{mailybaev2016spontaneously}, which occurs in the opposite limit $\Re\gg 1.$
This high-$\Re$ phenomenon has indeed features of an anomaly, but a more common name 
now is ``spontaneous stochasticity''. We discuss this next.

\subsection{Lagrangian spontaneous stochasticity} 

The immediate stimulus for the modern concept of spontaneous stochasticity 
was the experimental phenomenon of scalar anomalous dissipation in turbulent
flows. Consider a scalar such as a concentration field $c$ of a solute satisfying an advection-diffusion equation 
\be \partial_t c +\bu^\nu \bdot\grad c = D\triangle c. \lb{ceq} \ee 
When the advecting velocity field is incompressible ($\grad\bdot\bu^\nu=0$) 
and turbulent, then there is substantial empirical evidence that the scalar dissipation rate $\chi=D|\nabla c|^2$ scales as 
$\sim u^{\prime} c^{\prime\, 2}/L$ at high Reynolds number $\Re$ 
and high P\'eclet number $Pe=u^\prime L/D,$ becoming then independent both of the molecular viscosity $\nu$ and also of the molecular 
diffusivity $D$ \cite{donzis2005scalar}. This is the exact analogue for the 
``scalar intensity'' $I=(1/2)c^2$ of the dissipative anomaly in kinetic energy of the velocity field and it was assumed as a fundamental hypothesis in the dimensional 
mean-field theories of Obukhov \cite{obukhov1949structure} and Corrsin 
\cite{corrsin1951spectrum} for turbulent scalar advection. 

The seminal paper of Bernard, Gaw\c{e}dzki and Kupiainen \cite{bernard1998slow}  
(see also \cite{gawedzki1998intermittency}) studied such turbulent scalar 
advection using a Lagrangian path-integral representation. It is worth 
summarizing briefly their arguments. Equation \eqref{ceq} can be solved 
in integral form 
\be c(\bx,t) = \int d^dx_0\ c_0(\bx_0)\, p_{\nu,D}({\bf x}_0,0|{\bf x},t) \lb{soleq} 
\ee 
where $c_0$ is the initial data for the scalar field at time $t=0$ and 
$p_{\nu,D}({\bf x}_0,0|{\bf x},t)$ for $t>0$ is the transition probability backward 
in time of a stochastic Lagrangian tracer particle obeying 
\be d\bX = \bu^\nu(\bX,t) \, dt +\sqrt{2D} \, d{\bf W}(t). \lb{xeq} \ee 
In fact, for a scalar concentration field, the latter equation describes the 
evolution of a single solute molecule both advected by the velocity field $\bu^\nu$
and also subject to Brownian motion with molecular diffusivity $D.$ The transition 
probability in \eqref{soleq} has a standard Feynman-Ka\v{c} path-integral representation 
with an Onsager-Machlup action in the exponent 
\be p_{\nu,D}({\bf x}_0,0|{\bf x},t) = \frac{1}{\mathcal{Z}} \int_{{\bf X}(t)={\bf x}}
     \delta^d({\bf X}(0)-{\bf x}_0) \exp\left(-\frac{1}{4 D}\int_{0}^t d\tau\, |\dot{{\bf X}}(\tau)-{\bf u}^\nu({\bf X}(\tau),\tau)|^2\right)
     \mathcal{D}{\bf X}. \lb{pathint} \ee 
where the paths ${\bf X}$ start at ${\bf X}(t)=\bx$ and end at ${\bf X}(0)=\bx_0.$        
In a smooth, laminar flow, the above path-integral can be easily evaluated 
in the joint limit $\nu,D\to 0$ by a Laplace steepest-descent asymptotics as 
\be \lim_{D\to 0} p_{\nu,D}({\bf x}_0,0|{\bf x},t) = \delta^d(\bx_0-\bX(0;\bx,t))\  
\Longrightarrow \ c(\bx,t) = c_0(\bX(0;\bx,t))\ee 
where now $\bX$ solves the ODE for the limiting velocity $\bu$
\be \frac{d}{ds}\bX(s;\bx,t)=\bu(\bX(s;\bx,t),s),\ \quad  \bX(t;\bx,t)=\bx \lb{ODE} \ee 
and represents a standard deterministic Lagrangian fluid particle trajectory. This 
formal asymptotics is the rigorous result of Freidlin-Wentzell zero-noise 
large deviations theory for the SDE \eqref{xeq} when the velocity field $\bu$ 
obtained in the limit $\nu\to 0$ remains smooth. 

However, as noted earlier, Onsager had argued that for  a turbulent flow, the limiting velocity $\bu$ can remain at most 1/3 H\"older regular in the inviscid limit \cite{onsager1949statistical,eyink2006onsager,eyink2024onsager}. 
Bernard et al. \cite{bernard1998slow} then recalled 
the well-known result  \cite{hartman1982ordinary, agarwal1993uniqueness} 
that ODE's with velocities of such low regularity can have multiple, non-unique 
solutions $\bX(t)$ for exactly prescribed, deterministic initial data $\bX(0)=\bx_0$. 
Thus, there is no longer a unique action minimizer for the integrand in 
the representation \eqref{pathint} as $D,$ $\nu\to 0$ together. This led 
\cite{bernard1998slow} to conjecture that instead 
\be \lim_{\nu\to 0, Sc \hbox{ {\scriptsize fixed}}} p_{\nu,D}({\bf x}_0,0|{\bf x},t) = 
p_{*}({\bf x}_0,0|{\bf x},t) = \int d\mu_*(\omega)\, \delta^d(\bx_0-\bX^\omega(0;\bx,t)) \lb{LSpSt} \ee 
in the infinite Reynolds limit at fixed Schmidt number $Sc=\nu/D,$ where $\mu_*$
is some non-trivial (non-delta) measure over the set of unique solutions 
of the limiting ODE \eqref{ODE}. In fact, Bernard et al. \cite{bernard1998slow} 
presented convincing analytical evidence 
for this conclusion when the advecting velocity field $\bu^\nu$ is a fixed realization 
of a standard synthetic turbulence model, the Gaussian white-in-time ensemble of 
Kraichnan \cite{kraichnan1968small}. In that model, the origin of the non-uniqueness 
of limiting Lagrangian trajectories can be traced to the famous Richardson dispersion 
of particle pairs \cite{richardson1926atmospheric}. Richardson's scale-dependent 
diffusion equation for particle pair-separations in fact admits an exact self-similar 
solution, noted already by Richardson \cite{richardson1926atmospheric}, in which particles 
initially at the same location (zero separation) move apart to finite separations 
with probability one in any positive time. This long under-appreciated prediction of 
Richardson exactly underlies the conclusion \eqref{LSpSt} of  
Bernard et al. \cite{bernard1998slow} that Lagrangian particle 
trajectories become {\it spontaneously stochastic} in the joint limit $\Re,$ $Pe\to 0.$
In that case, the limiting concentration field satisfies 
\be c_*(\bx,t) = \lim_{\Re\to \infty, Sc \hbox{ {\scriptsize fixed}}} c_{\Re,Sc}(\bx,t) =
\int d^dx_0\ c_0(\bx_0)\, p_{*}({\bf x}_0,0|{\bf x},t).  
\ee 
and satisfies in distribution sense the ideal advection equation 
\be \partial_t c_*  +\bu \bdot\grad c_* = 0. \quad \ee
However, $p_{*}({\bf x}_0,0|{\bf x},t)$ is a non-trivial (non-delta) 
probability density that must satisfy 
$\int d^3x \   p_{*}({\bf x}_0,0|{\bf x},t)=1$ because of incompressibility of the 
advecting velocity field $\bu^*.$ Thus, by convexity of the quadratic function $f(c)=(1/2)c^2$
\be  \int d^3x \ \frac{1}{2}c_*^2({\bf x},t) < 
\int d^3x \ \int d^3x_0 \   p_{*}({\bf x}_0,0|{\bf x},t) \frac{1}{2}c_0^2({\bf x}_0)
=\int d^3x_0 \ \frac{1}{2}c_0^2({\bf x}_0)
\quad t>0 \ee 
In this way, \cite{bernard1998slow} explained the scalar dissipative anomaly as a consequence of the non-uniqueness and intrinsic randomness of Lagrangian 
particle paths in high Reynolds-number turbulent flows. 

Following works on the Kraichnan model have verified these conclusions and  
extended them in various ways, for example, to compressible versions of the model
\cite{gawedzki2000phase,E2000generalized}.  An authoritative 
review of the early results was written for physicists \cite{falkovich2001particles},
in which the term {\it spontaneous stochasticity} for the limiting behavior 
\eqref{LSpSt} first appeared, motivated by the physical analogy of remnant 
``spontaneous magnetization'' in ferromagnetic spin system with solutions remaining random even 
in the limit where governing equations become deterministic. 
Probabilists Le Jan \& Raimond have further proved all of these results as 
theorems for the Kraichnan model, as a rigorous application 
of Wiener chaos expansions \cite{lejan2002integration,lejan2004flows}. 
Corresponding results have since been found in related models of turbulent 
advection, both by physical arguments  
\cite{chaves2003lagrangian,peixoto2023spontaneous}
and by rigorous mathematical proofs 
\cite{armstrong2025anomalous,ruffenach2025spontaneous}, including 
for the active scalar described by Burgers' nonlinear diffusion equation 
\cite{eyink2015spontaneous} and recently for a passive scalar advected by a 
H\"older $1/3^-$ weak Euler solution constructed by convex integration 
techniques \cite{burczak2023anomalous}. 

For scalars advected by a physical turbulent flow that is governed by Navier-Stokes there is not yet a rigorous 
{\it a priori} demonstration of spontaneous stochasticity. However, Eyink \& Drivas 
\cite{drivas2017Alagrangian} have shown that Lagrangian spontaneous stochasticity 
is the only possible mechanism for anomalous scalar dissipation, for both passive 
and active scalars and regardless of the advecting incompressible velocity field, 
at least away from solid walls \cite{drivas2017Blagrangian}. In fact, the 
simple argument of \cite{bernard1998slow} for a decaying 
scalar sketched above can be easily turned into a rigorous proof, but additional methods 
of \cite{drivas2017Alagrangian} establish the necessity of spontaneous stochasticity 
for even more general situations such as statistical steady states with constant 
injection of scalar intensity. Thus, the known empirical evidence for scalar 
anomalous dissipation \cite{donzis2005scalar} provides indirect empirical support 
for spontaneous stochasticity. The simplest direct evidence would be observation 
of the fundamental Lagrangian mechanism of the ``forgetting'' of initial 
separations or small random perturbations on Lagrangian particle trajectories 
evolved for a finite fraction of the large-eddy turnover time $T=L/u^\prime.$ 
This effect, in the limit of large Reynolds numbers, requires a very long time 
measured in Kolmogorov dissipation units. Numerical simulations of high Reynolds-number 
turbulence supply robust evidence of such ``forgetting'', both of molecular diffusivity of stochastic Lagrangian trajectories \cite{eyink2011stochastic,buaria2016lagrangian} and of initial separations of 
deterministic Lagrangian trajectories \cite{bitane2013geometry,buaria2015characteristics}. 
Laboratory experiments can easily reach much higher Reynolds numbers but it is very difficult to track for a finite fraction of the large eddy turnover time sufficiently many particles 
initially close together. Despite intensive efforts \cite{bourgoin2006role,ouellette2006experimental,tan2022universality}, the evidence 
from laboratory experiment remains inconclusive. 

Since the basic physical mechanism of Lagrangian spontaneous stochasticity is 
``forgetting'' initial data  and/or small random perturbations at long microscopic 
times, the renormalization group (RG) method should be applicable to determine the possible 
universality classes of spontaneous statistics. There is an obvious close analogy 
to the ``forgetting'' of the detailed microscopic Hamiltonian in equilibrium critical 
phenomena, with broad universality classes exhibiting the same universal scaling behavior 
at large distances near the critical point. Furthermore, many common fluid mechanical models possess 
space-time scaling symmetries, so that the macroscopic weak-noise limit is mathematically 
equivalent to the microscopic long-time limit. This suggests that RG methods 
can be applied to analyze spontaneous stochasticity that were developed to establish self-similar 
or front solutions to PDE's and their domains of attraction in the long-time limit 
\cite{goldenfeld1991asymptotics,chen1994selection,bricmont1995renormalizing,bricmont1999stability}. 
An example of such an RG analysis was carried out for a ``minimal ODE model'' of spontaneous stochasticity \cite{eyink2020renormalization}, which permitted a complete classification of all possible fixed points, 
an exact characterization of their domains of attraction, and a description of the approach to 
each fixed point. The latter long-time RG result corresponds to a novel large-deviations asymptotics 
for the transition probability \eqref{pathint} in the singular, vanishing-noise limit $\Re,$ $Pe\to\infty,$ distinct from the standard Freidlin-Wentzell result and mediated by non-unique solutions of the 
limiting singular ODE rather than by usual instanton trajectories.

An important result of the RG analysis in \cite{eyink2020renormalization} 
was a phase diagram in the $\ln(\Re)$-$\ln(Pe)$-plane, taking as order 
parameter the variance of the particle position $X$ at unit macroscopic time. 
Although there are no perfectly sharp transition boundaries, three clearly defined 
phases emerged: (i) a {\it noise-dominated phase} for $Pe\lesssim Pe_c$ with large
variance dependent upon $D$ (ii) a {\it deterministic phase} for $Sc=Pe/\Re\gtrsim \exp(\sqrt{\Re})$
with variance nearly zero, and (iii) the {\it spontaneously stochastic phase} in the 
rest of parameter space with the same variance as reached in the joint limit $\Re,$ $Pe\to \infty.$
Thus, even though the critical point lies formally at $\Re=Pe=\infty,$ which 
is strictly unattainable, the spontaneous statistics are observed 
for a large swathe of the phase diagram and can be observed in physical experiments 
at sufficiently large, but finite, $\Re$ and $Pe.$ This situation is reminiscent of quantum critical 
systems, for example, where the critical point occurs at zero absolute temperature 
and is thus similarly unattainable (Nernst law). Nevertheless, the influence of this 
critical point expands to a broad nonzero temperature regime of quantum criticality, 
enabling its study through a variety of experimental techniques. The situation with 
spontaneous stochasticity is quite analogous. 

\subsection{Eulerian spontaneous stochasticity} 

Since the essential ingredient for spontaneous stochasticity is a multiplicity 
of solutions to the Cauchy problem and since dissipative,  
H\"older-continuous solutions of ideal Euler equations are proved to be 
non-unique \cite{daneri2021non,delellis2023nonuniqueness,isett2022nonuniqueness,giri2023L3},
it is natural to wonder whether the entire velocity field might become spontaneously 
stochastic in the high-Reynolds-number limit, and not just Lagrangian particle 
trajectories \cite{eyink2014mathematical}. 
In fact, this possibility was anticipated in a remarkable
1969 paper of Edward Lorenz \cite{lorenz1969predictability}, the first sentence 
of whose abstract reads as follows: 
\begin{quotation}
\noindent 
``It is proposed that certain formally deterministic fluid systems which possess many scales of motion are observationally indistinguishable from indeterministic systems; specifically, that two states of the system differing initially by a small “observational error” will evolve into two states differing as greatly as randomly chosen states of the system within a finite time interval, which cannot be lengthened by reducing the amplitude of the initial error.''
\end{quotation}
The concordances with spontaneous stochasticity are striking. In particular, Lorenz 
argues that ``formally deterministic'' equations may exhibit solutions with dynamics 
``indistinguishable from indeterministic''. Our limit of high Reynolds number is his 
``many scales of motion'' and our limit of vanishing noise is his 
``reducing the amplitude of the initial error,''
with the solution remaining as stochastic as ``randomly chosen states'' in the limit. 
Furthermore, Lorenz proposed a specific mechanism leading to these results, 
sometimes now called the ``inverse error cascade'', in which tiny randomness at small 
spatial scales would progress stepwise up to larger scales. This inverse error cascade process 
has been subsequently well-verified in high Reynolds-number turbulent flows both by spectral closure 
calculations \cite{leith1972predictability} and by direct numerical simulations 
\cite{boffetta2001predictability,boffetta2017chaos}. 

It seems that Lorenz' 1969 paper, when recalled at all, has been widely misinterpreted 
both by physicists and by mathematicians to be an application of deterministic 
chaos as expounded in the more famous 1963 paper \cite{lorenz1963deterministic}. 
Lorenz,  however, took great pains himself in the later 1969 work to distinguish the 
two phenomena, emphasizing in the second paper that the time-interval of predictability 
in that case ``cannot be lengthened by reducing the amplitude of the initial error.''
This is quite distinct from ordinary chaos where the time-horizon of predictability 
can be made arbitrarily long, in principle, by decreasing initial error. In fact, the 
persistent stochasticity over any finite time interval with vanishing initial error
requires a Lyapunov exponent not merely positive but even diverging to infinity. Physicist and 
climatologist T. Palmer has made a compelling case \cite{palmer2014real,palmer2024real} 
that Lorenz himself used the term ``butterfly effect'' to indicate the phenomenon 
of his 1969 paper and not deterministic chaos. Lorenz' ``real butter-fly effect'' 
\cite{lorenz1969predictability} was apparently first identified with the spontaneous 
stochasticity phenomenon of Bernard et al. \cite{bernard1998slow} by A. Mailybaev \cite{mailybaev2016spontaneously}, who showed by numerical simulations that both phenomena 
coincide in a shell model of turbulence. Lorenz had in fact foreseen almost all 
of the modern concept except for one crucial element that was clearly identified by 
Bernard et al. \cite{bernard1998slow}: the requisite non-uniqueness of solutions
of the limiting ``formally deterministic'' (but singular) evolution equation, even 
with precisely specified, deterministic initial data. This fundamental insight makes 
clear the precise mathematical formulation of Lorenz' proposals in his 1969 
paper and highlights their revolutionary departure from conventional theory of 
ODE's and PDE's at the time.

Since the first computational work of Mailybaev on Eulerian spontaneous stochasticity 
in shell models \cite{mailybaev2016spontaneously}, the phenomenon has been found 
to occur in a number of other situations. First, several mathematical toy models 
have been identified in which it is possible to prove Eulerian spontaneous stochasticity 
rigorously \cite{drivas2021life,drivas2024statistical,mailybaev2023spontaneously}.
The ``spontaneously stochastic Arnold's cat'' model \cite{mailybaev2023spontaneously}
makes especially transparent the role of multi-scale chaotic dynamics with time scales 
(inverse Lyapunov exponents) decreasing rapidly to zero at small length scales, a feature 
observed numerically also in shell models \cite{yamada1998asymptotic}. Perhaps most 
impressive is the numerical evidence of Eulerian spontaneous stochasticity for 
realistic fluid equations with singular initial data subject to classical linear 
instabilities, especially the Kelvin-Helmholtz unstable vortex sheet \cite{thalabard2020butterfly}
and a sharp density interface subject to Rayleigh-Taylor instability \cite{mailybaev2017toward,biferale2018rayleigh}. Note that the required non-uniqueness
of admissible weak solutions to the ideal limit equations has been proved for both 
of these singular initial data, Kelvin-Helmholtz \cite{szekelyhidi2011weak,mengual2023dissipative}
and Rayleigh-Taylor \cite{gebhard2021new,gebhard2025rayleigh}. The ubiquity of these 
prototypical instabilities in fluid flows in engineering, geophysics, and astrophysics 
suggests that Eulerian spontaneous stochasticity may be a quite common occurence. Indeed, 
when blowing out a candle, who expects the plume to billow the same way each time? 

Although spontaneous stochasticity may be prompted by any vanishingly small random
perturbation, it is especially interesting for fundamental statistical physics to inquire 
whether it may be triggered by thermal noise due to molecular agitation. Pioneering work 
was done in this area almost 70 years ago by R. Betchov \cite{betchov1957fine,betchov1961thermal},
who predicted that thermal noise would strongly modify the dissipation range of turbulent flows.
Furthermore, Betchov argued that linear instabilities could amplify the ``steady supply of 
completely random fluctuations, emerging out of the kinetic noise'' and he went on to argue that 
\begin{quotation}
\noindent 
``For larger eddy Reynolds numbers, the amplification required for a similar effect is much larger but since it may be furnished in several stages, the basic idea can still be used. Thus, the cascade of energy from large eddies to small eddies is perhaps associated with the constant build up of unpredictable new eddies, out of the thermal agitation.''
\end{quotation}
Betchov's prescient insights on thermal noise effects in the dissipation range were rediscovered by Bandak et al. \cite{bandak2022dissipation} who also checked them in a shell-model simulation. Shortly thereafter,
the conclusions were verified in a moderate Reynolds number turbulent simulation using the full nonlinear Landau-Lifschitz fluctuating hydrodynamics equations by J. Bell et al. \cite{bell_nonaka_garcia_eyink_2022}. 
The question remained, however, whether thermal noise might trigger inverse error cascade and Eulerian spontaneous stochasticity. As previously discussed, there is expected to be a minimum level of noise,
dependent on Reynolds number, that is sufficient for this purpose and it is not entirely obvious 
that tiny thermal noise effects suffice. 

To make the issues more concrete, consider the standard incompressible, low Mach-number version of the fluctuating hydrodynamics equations \cite{forster1977large,donev2014low}: 
\be \partial_t{\bf u} + P_\Lambda({\bf u}\cdot\boldsymbol{\nabla}){\bf u} = -\boldsymbol{\nabla} p + \nu\Delta {\bf u} + \sqrt{\frac{2\nu k_BT}{\rho}} \boldsymbol{\nabla}\cdot \boldsymbol{\eta}+{\bf f},\qquad \boldsymbol{\nabla\cdot}{\bf u}=0 \lb{FNS} \ee 
with the Gaussian noise $\boldsymbol{\eta}$ white-in-time with zero mean and covariance
\be \langle \eta_{ij}({\bf x},t) \eta_{kl}({\bf x}',t')\rangle = 
\left(\delta_{ik}\delta_{jl}+\delta_{il}\delta_{jk} -\frac{2}{3}\delta_{ij}\delta_{kl}\right) \delta^3_\Lambda({\bf x}-{\bf x}')\delta(t-t'). \lb{FDT} \ee 
This form of noise is, of course, mandated by the fluctuation-dissipation relation. Note that these 
equations are valid only with a high-wavenumber (UV) cut-off $\Lambda,$ which must be chosen 
so that $\Lambda^{-1}$ lies between the molecular mean-free-path length and the hydrodynamic 
gradient-length (roughly the Kolmogorov scale in a turbulent flow). We thus interpret the 
equations \eqref{FNS} in the sense of an ``effective field theory'' with fluid parameters 
such as kinematic viscosity $\nu_\Lambda$ explicitly $\Lambda$-dependent and governed 
by RG flow equations \cite{forster1977large}. Thus, the spectral projection $P_\Lambda$
projects the quadratic nonlinear term back to wavenumbers with magnitude $<\Lambda$
and the noise term contains the approximate delta function $\delta_\Lambda(\bx)
=\frac{1}{L^3}\sum_{|\bk|<\Lambda} e^{i\bk\bdot\bx}$ in the periodic box of size $L.$
Non-dimensionalized with large-scale quantities $L$ and $U$ (say, the rms velocity), 
then the governing equations become
\be \partial_t{\bf u} + P_{\hat{\Lambda}}({\bf u}\cdot\boldsymbol{\nabla}){\bf u} = -\boldsymbol{\nabla} p + \frac{1}{\Re} \Delta {\bf u} + \frac{1}{\Re^{15/8}}\sqrt{2\theta_K} \boldsymbol{\nabla}\cdot \boldsymbol{\eta} + \digamma_L {\bf f} \lb{FNS-non} \ee
with characteristic dimensionless number groups 
\be \Re=\frac{UL}{\nu}, \qquad 
\theta_\eta= \frac{k_B T}{\rho \nu^{11/4}\varepsilon^{-1/4} }, \qquad
\digamma_L=\frac{f_{rms}L}{U^2}, \qquad
\hat{\Lambda}=\Lambda L\in [\Re^{3/4},\Re/Ma] \ee 
the Reynolds number, the thermal energy relative to kinetic energy of a Kolmogorov-scale eddy,
the inefficiency of the deterministic body force ${\bf f}$ and the non-dimensional
wavenumber cut-off, respectively. Note that we have assumed here the Taylor scaling for energy dissipation,  $\varepsilon \sim U^3/L,$ and $\theta_\eta,$ $\digamma_L$ are expected to be constant (or nearly so) in the 
limit $\Re\to \infty,$ while $\hat{\Lambda}\to\infty.$ The latter result, by the way, makes clear 
that the ``continuum approximation'' is an idealized description that applies only in the 
limit $\Re\to\infty.$ We may now take a deterministic initial condition $\bu_0$ at time 
$t_0$ and consider the transition probability $P_{\Re}[\bu,t|\bu_0,t_0]$ at time $t.$ 
This probability density may be written, in the same way as for scalar advection, as a functional path-integral 
\be P_{\Re}[\bu,t|\bu_0,t_0]
= \int_{{\bf v}(0)={\bf u}_0}
     \delta[{\bf v}(t)-{\bf u}] \exp\left(-S_{\Re}[\bv]\right)
     \mathcal{D}{\bf v}. \lb{transP-FNS} \ee 
with Onsager-Machlup action now given by 
\be S_{\Re}[\bv]=\frac{\Re^{15/4}}{4 \theta_\eta}\int_{t_0}^t d\tau\, \left\|
\dot{{\bf v}}+ + P_{\hat{\Lambda}}({\bf v}\cdot\boldsymbol{\nabla}){\bf v} +\boldsymbol{\nabla} p 
- \frac{1}{\Re} \Delta {\bf v} -\digamma_L {\bf f}\right\|_{-1}^2 \lb{pathint-FNS} \ee 
where $\|\bF\|_{-1}^2:=\langle\bF,(-\triangle)\bF\rangle.$ Exactly as in the Lagrangian case, 
this path-integral would be na\"ively evaluated in the limit $\Re\to\infty$ by steepest descent method,
leading to the conclusion that the dynamics is deterministic in that limit and described by 
the limiting solution of the incompressible Euler equations. However, these solutions 
are generally non-unique, even for fixed initial data $\bu_0,$ so that spontaneous 
stochasticity can be anticipated here. 

It is impossible with existing numerical methods and computational resources to 
simulate the model \eqref{FNS} (or, non-dimensionalized, \eqref{FNS-non}) for $\Re\gg 1.$
Thus, the problem has been tackled instead with simplifications that permit computations 
at sufficiently high Reynolds numbers. First, Bandak et al. \cite{bandak2024spontaneous} 
used a Sabra shell-model version of the dynamics \eqref{FNS}. To avoid waiting for smooth 
initial data to develop singularities by finite-time blow-up, \cite{bandak2024spontaneous}
considered two different singular initial data, an exact steady (but unstable) K41 solution 
of the inviscid limit equations with H\"older exponent $h=\frac{1}{3}$ and an initial data
selected from a very high-$\Re$ turbulent cascade state of the deterministic Sabra model 
with $h\doteq 0.2.$ More recently, Ortiz et al. \cite{ortiz2025spontaneous} have instead 
simulated \eqref{FNS} with a logarithmic lattice approximation. They chose two power-law
initial data, one rough with  $h=\frac{1}{2}$ and another smooth with  $h=\frac{3}{2}.$
Thus, in the latter case, the infinite-$\Re$ limit is guaranteed to be deterministic 
until after the blow-up time. Both numerical studies obtain results consistent 
with the hypothesis that the transition probability \eqref{transP-FNS} is nontrivial 
in the limit $\Re\to\infty,$ within the model approximations. The early stages of the growth 
of velocity variance are Brownian-like and driven just by the thermal noise. However, 
in a time of order the Kolmogorov dissipation time $t_\eta=(\nu/\varepsilon)^{1/2}$
this initial perturbation becomes amplified by small-scale chaotic dynamics, as 
first proposed by Ruelle \cite{ruelle1979microscopic}. Once the perturbation grows 
to the size of the smallest-scale inertial eddy, the nonlinear inverse error cascade of Lorenz\cite{lorenz1969predictability} takes over and propagates the randomness through 
the inertial range, reaching the length-scale $\ell$ in the eddy-turnover time 
for the initial data at that scale, with larger-scale eddies essentially ``frozen.'' 
Mathematically, it can be proved \cite{eyink2025space} that the transition probability 
may remain non-trivial as $\Re\to \infty$ only if the limiting ``formally deterministic'' 
inviscid equations have non-unique weak solutions for the given fixed initial data. 

Just as for the Lagrangian case, renormalization group methods have been developed 
to analyze Eulerian spontaneous stochasticity, primarily by Mailybaev and collaborators 
\cite{mailybaev2023spontaneous,mailybaev2025rgA,mailybaev2025rgB,mailybaev2025rgC}.
These methods exploit also the space-time scale symmetries of the limiting inviscid 
dynamics, but they have more of the dynamical systems flavor of the Feigenbaum RG 
for flow maps. Without going into full details, the RG transformation maps a regularized 
form of the inviscid dynamics to a similar model dynamics with reduced regularization.
Thus, RG fixed points are directly associated with the singular, inviscid limit dynamics.
It is still not clear how the different RG approaches are related to each other 
and what is their specific domains of applicability. For example, the Ph.D. thesis 
of Peng \cite{peng2025spontaneous} implements in the "spontaneous stochastic Arnold's cat"
model a version of the long-time RG previously described for Lagrangian spontaneous stochasticity.
He finds by mathematically rigorous analysis an RG fixed point which provides a 
universal reduced stochastic model of the modes at long times and large scales. 
The RG method recently elaborated for Eulerian spontaneous stochasticity in 
shell models \cite{mailybaev2025rgC} likewise features stochastic large-eddy 
simulation models. Clarifying the relations of these different RG approaches is 
important work for the future. 

Spontaneous stochasticity is a novel form of randomness in turbulence theory. Traditional  
probabilistic approaches to turbulence have assumed random ensembles of initial data
and/or random forces or else appealed to presumed ergodicity to replace such ensemble
averages with time- and/or space-averages. The randomness associated to spontaneous 
stochasticity appears for fixed realizations of initial data and forcing
and is associated instead to the non-uniqueness of weak solutions to singular 
ODE's \cite{hartman1982ordinary, agarwal1993uniqueness} and PDE's 
\cite{daneri2021non,isett2022nonuniqueness,delellis2023nonuniqueness,giri2023L3}
that arise in the ``inviscid'' or infinite Reynolds number limit. The path-integral 
expressions \eqref{pathint} and \eqref{pathint-FNS} make clear the analogy of 
spontaneous stochasticity to equilibrium statistical-mechanical spin systems 
with multiple zero-temperature ground states. Minimizers of the limiting 
Onsager-Machlup action (Euler solutions) play the role of ground states and 
Reynolds number plays the role of inverse temperature. As already emphasized 
by us decades ago \cite{eyink1994analogies}, the infinite Reynolds number limit 
corresponds to approaching a critical point but, even more precisely, a 
zero-temperature critical point. Thus, spontaneous stochasticity
falls into the framework of ``multiple equilibria'', which has been emphasized by 
Parisi \cite{parisi2023nobel} to be a central paradigm in modern physics 
of complex systems. 

\section{Discussion and synthesis}
\label{sec:disc}

The main theme of this article has been the emerging understanding of what are the intrinsic statistical properties that arise in turbulent fluids.  We have seen that there are two regimes --- near the onset of turbulence and at fully-developed turbulence --- where universal phenomena are found, but for different reasons.  We conclude with a summary, and some remarks about unsolved problems.

The onset of turbulence in sub-critical transitions generates the statistical behaviour of directed percolation, but  this has been understood mechanistically only in quasi-one-dimensional flows, but not yet in two-dimensional flows.  Even though the fluctuations and scaling laws can be calculated and measured, the question of how stochasticity is generated is subtle.  

Central to our understanding of the behaviour at asymptotically large Reynolds numbers, is that, depending on the flow and its boundary conditions, turbulence exhibits a dissipation rate that is anomalous in one of two ways: (i) it  becomes independent of Reynolds number (strong anomaly), or (ii) the corresponding drag or friction factor decays more slowly with Reynolds number than it would in the laminar case (weak anomaly).  The phenomenon is termed anomalous, because if one asked what is the governing equation for the fluid flow in the limit of asymptotically large Reynolds number, the answer would seem to be the Euler equations, which conserve kinetic energy.  Thus there would naively be no dissipation at all.  The misunderstanding is that one cannot set $\Re=\infty$ in the limiting process.  Where does the energy go to in the limit of $\Re\rightarrow\infty$?  For the problem to be well-defined, the macroscopic velocity field needs to be defined with respect to an ultraviolet cutoff (e.g. Onsager's point-splitting or a coarse-graining length), and the governing equations are actually an effective field theory for the coarse-grained velocity field.  The degrees of freedom on scales smaller than the cutoff are not resolved and are in effect stochastic noise.  The energy from large scales flows through the turbulent fluid into these sub-resolution scales, and generates Joule heating, even in the limit of $\Re\rightarrow\infty$.

The other development has been the recognition of a strong form of intrinsic randomness in models of turbulent flow, now known as spontaneous stochasticity.  In contrast to chaotic dynamics, whereby Lagrangian trajectories of fluid parcels deviate exponentially in time, Lagrangian spontaneous stochasticity allows trajectories to be arbitrarily far apart in a finite time, and this separation cannot be reduced by making the initial separation smaller!  Contrary to some statements in the literature, spontaneous stochasticity is not an infinite Reynolds number phenomenon only.  It arises at asymptotically large Reynolds numbers, and reflects the influence of a singular point as $\Re\rightarrow\infty$.  This statement is not equivalent to saying that spontaneous stochasticity is the outcome of taking the $\Re=\infty$ limit of the Navier-Stokes equations, and is thus purely a property of the Euler equations. This has been demonstrated conclusively in the Sabra shell model of turbulence, for example \cite{bandak2024spontaneous}.

Spontaneous stochasticity and dissipative anomalies have certain features in common.  Clearly, both are properties of asymptotically large Reynolds number, governed by the Navier-Stokes equations, but reflecting aspects of the Euler equations.  But exactly how are they related?  For example, does the presence of spontaneous stochasticity necessarily imply a dissipative anomaly, and if so, which type?  Can one have a dissipative anomaly but not spontaneous stochasticity?  The answers to these questions are beginning to be understood.  For example, it is known that for any advected scalar (active or passive), the existence of a dissipative anomaly requires Lagrangian spontaneous stochasticity \cite{drivas2017Alagrangian}.  On the other hand, there are situations where a dissipative anomaly exists, but there is no Eulerian spontaneous stochasticity.  This arises in Burgers equation \cite{eyink2015spontaneous} and the passive scalar advection of the Kraichnan model \cite{lototskii2004passive}.   

Another question is to understand whether the dissipative anomaly and spontaneous stochasticity have a common origin.  We speculate that this is possible, and is associated with the boundaries of the flow, specifically the stochastic separation 
of vortex sheets from the wall.  This would be consistent with the experimental findings in the Nikuradse experiments on pipe flow, where the strong dissipative anomaly for the friction factor has a magnitude that is determined by the wall roughness.

Beyond fundamental issues of statistical physics, spontaneous stochasticity deeply impacts 
research problems in many other areas of physics. Lagrangian spontaneous stochasticity has 
been suggested as an explanation for the ultimate regime in Rayleigh-B\'enard turbulence \cite{eyink2018lagrangian}, for fast magnetic reconnection in astrophysics and space science \cite{lazarian1999reconnection,eyink2011fast,eyink2013flux}, for quantized vortex dynamics  
in superfluid turbulence \cite{tang2025turbulent}, for entropy cascade in collisionless 
plasma turbulence \cite{krommes1994role,schekochihin2009astrophysical,bardos2020onsager}, 
and for the semiclassical limit of singular quantum dynamics \cite{athanassoulis2012strong,eyink2015quantum}. Eulerian spontaneous stochasticity 
was already understood in the pioneering work of Lorenz \cite{lorenz1969predictability}
to imply essential limits of predictability in weather and climate systems. 
Renormalization group approaches to spontaneous stochasticity 
\cite{peng2025spontaneous,mailybaev2025rgC} make clear its deep connections with 
stochastic climate modeling \cite{hasselmann1976stochastic,rose1977eddy,leith1990stochastic,eyink1996turbulence,palmer2019stochastic},
providing both a mathematical and physical foundation for such models and also a 
methodology for deriving stochastic models in a systematic manner. The implications
of intrinsic predictability extend also to areas such as astrophysics \cite{genel2019quantification}
and cosmology \cite{neyrinck2022boundaries} which traditionally have been thought to 
exemplify the paradigm of Laplacian determinism. Spontaneous stochasticity brings to completion 
the revolution begun by chaos theory, so that the proper goal in classical dynamics, just as 
in quantum dynamics, is not to predict a certain future but instead to forecast ensemble probabilities. 

\bigskip

\enlargethispage{20pt}


\dataccess{No data were generated in the preparation of this paper.}

\aucontribute{Both authors conceived and wrote the manuscript.}

\competing{The author(s) declare that they have no competing interests.}

\funding{This work was partially supported by the Simons Foundation through Targeted Grant “Revisiting the Turbulence Problem Using Statistical Mechanics” [Grants 662985 (NG) and 663054 (GE)] and the Simons Collaboration on Wave Turbulence [Grant 1151717 (NG) and  (GE)].}

\ack{The authors gratefully acknowledge many discussions on the topics of this paper with our collaborators and colleagues on some of the work surveyed: A. Mailybaev, D. Bandak, H.-Y. Shih, X. Wang, B. Hof, D. Barkley, H. Quan, S. Kumar, L. Peng, T. D. Drivas, K. Iyer, and K. R. Sreenivasan}



\bibliographystyle{ieeetr}
\bibliography{bibliography}

\begin{thebibliography}{100}

\bibitem{sreenivasan1986transition}
K.~Sreenivasan and R.~Ramshankar, ``Transition intermittency in open flows, and
  intermittency routes to chaos,'' {\em Physica D: Nonlinear Phenomena},
  vol.~23, no.~1-3, pp.~246--258, 1986.

\bibitem{crutchfield1988attractors}
J.~P. Crutchfield and K.~Kaneko, ``Are attractors relevant to turbulence?,''
  {\em Physical Review Letters}, vol.~60, pp.~2715--2718, 1988.

\bibitem{pomeau}
Y.~Pomeau, ``Front motion, metastability and subcritical bifurcations in
  hydrodynamics,'' {\em Physica}, vol.~23D, pp.~3--11, 1986.

\bibitem{manneville2016transition}
P.~Manneville, ``Transition to turbulence in wall-bounded flows: Where do we
  stand?,'' {\em Mechanical Engineering Reviews}, vol.~3, pp.~15--00684, 2016.

\bibitem{barkley2016theoretical}
D.~Barkley, ``Theoretical perspective on the route to turbulence in a pipe,''
  {\em J. Fluid Mech}, vol.~803, no.~1, 2016.

\bibitem{goldenfeld2017statistical}
N.~Goldenfeld, ``A statistical mechanical phase transition to turbulence in a
  model shear flow,'' {\em Journal of Fluid Mechanics}, vol.~830, pp.~1--4,
  2017.

\bibitem{goldenfeld2017turbulence}
N.~Goldenfeld and H.-Y. Shih, ``Turbulence as a problem in non-equilibrium
  statistical mechanics,'' {\em Journal of Statistical Physics}, vol.~167,
  no.~3-4, pp.~575--594, 2017.

\bibitem{eckhardt2018transition}
B.~Eckhardt, ``Transition to turbulence in shear flows,'' {\em Physica A:
  Statistical Mechanics and its Applications}, vol.~504, pp.~121--129, 2018.

\bibitem{mukund2018critical}
V.~Mukund and B.~Hof, ``The critical point of the transition to turbulence in
  pipe flow,'' {\em Journal of Fluid Mechanics}, vol.~839, pp.~76--94, 2018.

\bibitem{avila2023transition}
M.~Avila, D.~Barkley, and B.~Hof, ``Transition to turbulence in pipe flow,''
  {\em Annual Review of Fluid Mechanics}, vol.~55, pp.~575--602, 2023.

\bibitem{hof2008repeller}
B.~Hof, A.~de~Lozar, D.~J. Kuik, and J.~Westerweel, ``Repeller or attractor?
  selecting the dynamical model for the onset of turbulence in pipe flow,''
  {\em Physical Review Letters}, vol.~101, no.~21, p.~214501, 2008.

\bibitem{goldenfeld2010extreme}
N.~Goldenfeld, N.~Guttenberg, and G.~Gioia, ``Extreme fluctuations and the
  finite lifetime of the turbulent state,'' {\em Physical Review E}, vol.~81,
  no.~3, p.~035304, 2010.

\bibitem{avila2011onset}
K.~Avila, D.~Moxey, A.~de~Lozar, M.~Avila, D.~Barkley, and B.~Hof, ``The onset
  of turbulence in pipe flow,'' {\em Science}, vol.~333, no.~6039,
  pp.~192--196, 2011.

\bibitem{barkley2011simplifying}
D.~Barkley, ``Simplifying the complexity of pipe flow,'' {\em Physical Review
  E}, vol.~84, no.~1, p.~016309, 2011.

\bibitem{sipos2011directed}
M.~Sipos and N.~Goldenfeld, ``Directed percolation describes lifetime and
  growth of turbulent puffs and slugs,'' {\em Physical Review E}, vol.~84,
  no.~3, p.~035304, 2011.

\bibitem{barkley2012pipe}
D.~Barkley, ``Pipe flow as an excitable medium,'' {\em Revista Cubana de
  F{\'i}sica}, vol.~29, no.~1E, pp.~1--27, 2012.

\bibitem{song2014deterministic}
B.~Song and B.~Hof, ``Deterministic and stochastic aspects of the transition to
  turbulence,'' {\em Journal of Statistical Mechanics: Theory and Experiment},
  vol.~2014, no.~2, p.~P02001, 2014.

\bibitem{barkley2015rise}
D.~Barkley, B.~Song, V.~Mukund, G.~Lemoult, M.~Avila, and B.~Hof, ``The rise of
  fully turbulent flow,'' {\em Nature}, vol.~526, no.~7574, pp.~550--553, 2015.

\bibitem{hof_dp_1d_couette}
G.~Lemoult, L.~Shi, K.~Avila, S.~V. Jalikop, M.~Avila, and B.~Hof, ``Directed
  percolation phase transition to sustained turbulence in couette flow,'' {\em
  Nature Physics}, vol.~12, no.~3, pp.~254--258, 2016.

\bibitem{ppmodel}
H.-Y. Shih, T.-L. Hsieh, and N.~Goldenfeld, ``Ecological collapse and the
  emergence of travelling waves at the onset of shear turbulence,'' {\em Nature
  Physics}, vol.~12, no.~3, pp.~245--248, 2016.

\bibitem{chantry2017universal}
M.~Chantry, L.~S. Tuckerman, and D.~Barkley, ``Universal continuous transition
  to turbulence in a planar shear flow,'' {\em Journal of Fluid Mechanics},
  vol.~824, 2017.

\bibitem{budanur2020upper}
N.~B. Budanur, E.~Marensi, A.~P. Willis, and B.~Hof, ``Upper edge of chaos and
  the energetics of transition in pipe flow,'' {\em Physical Review Fluids},
  vol.~5, no.~2, p.~023903, 2020.

\bibitem{gome2020statistical}
S.~Gom{\'e}, L.~S. Tuckerman, and D.~Barkley, ``Statistical transition to
  turbulence in plane channel flow,'' {\em Physical Review Fluids}, vol.~5,
  no.~8, p.~083905, 2020.

\bibitem{wang2022stochastic}
X.~Wang, H.-Y. Shih, and N.~Goldenfeld, ``Stochastic model for
  quasi-one-dimensional transitional turbulence with streamwise shear
  interactions,'' {\em Physical Review Letters}, vol.~129, no.~3, p.~034501,
  2022.

\bibitem{klotz2022phase}
L.~Klotz, G.~Lemoult, K.~Avila, and B.~Hof, ``Phase transition to turbulence in
  spatially extended shear flows,'' {\em Physical Review Letters}, vol.~128,
  no.~1, p.~014502, 2022.

\bibitem{hof2023directed}
B.~Hof, ``Directed percolation and the transition to turbulence,'' {\em Nature
  Reviews Physics}, vol.~5, no.~1, pp.~62--72, 2023.

\bibitem{lemoult2024directed}
G.~Lemoult, V.~Mukund, H.-Y. Shih, G.~Linga, J.~Mathiesen, N.~Goldenfeld, and
  B.~Hof, ``Directed percolation and puff jamming near the transition to pipe
  turbulence,'' {\em Nature Physics}, pp.~1--7, 2024.

\bibitem{hinrichsen2000non}
H.~Hinrichsen, ``Non-equilibrium critical phenomena and phase transitions into
  absorbing states,'' {\em Advances in physics}, vol.~49, no.~7, pp.~815--958,
  2000.

\bibitem{lorenz1969predictability}
E.~N. Lorenz, ``The predictability of a flow which possesses many scales of
  motion,'' {\em Tellus}, vol.~21, no.~3, pp.~289--307, 1969.

\bibitem{bernard1998slow}
D.~Bernard, K.~Gawedzki, and A.~Kupiainen, ``Slow modes in passive advection,''
  {\em Journal of Statistical Physics}, vol.~90, no.~3, pp.~519--569, 1998.

\bibitem{gawedzki2000phase}
K.~Gaw\c{e}dzki and M.~Vergassola, ``Phase transition in the passive scalar
  advection,'' {\em Physica D: Nonlinear Phenomena}, vol.~138, no.~1-2,
  pp.~63--90, 2000.

\bibitem{E2000generalized}
W.~E and E.~vanden Eijnden, ``Generalized flows, intrinsic stochasticity, and
  turbulent transport,'' {\em Proceedings of the National Academy of Sciences},
  vol.~97, no.~15, pp.~8200--8205, 2000.

\bibitem{lejan2002integration}
Y.~Le~Jan and O.~Raimond, ``Integration of {B}rownian vector fields,'' {\em
  Ann. Probab.}, vol.~30, no.~2, pp.~826--873, 2002.

\bibitem{lejan2004flows}
Y.~Le~Jan and O.~Raimond, ``Flows, coalescence and noise,'' {\em Ann. Probab.},
  vol.~32, no.~2, pp.~1247--1315, 2004.

\bibitem{palmer2014real}
T.~Palmer, A.~D{\"o}ring, and G.~Seregin, ``The real butterfly effect,'' {\em
  Nonlinearity}, vol.~27, no.~9, p.~R123, 2014.

\bibitem{eyink2015spontaneous}
G.~L. Eyink and T.~D. Drivas, ``Spontaneous stochasticity and anomalous
  dissipation for burgers equation,'' {\em Journal of Statistical Physics},
  vol.~158, no.~2, pp.~386--432, 2015.

\bibitem{mailybaev2016spontaneously}
A.~A. Mailybaev, ``Spontaneously stochastic solutions in one-dimensional
  inviscid systems,'' {\em Nonlinearity}, vol.~29, no.~8, p.~2238, 2016.

\bibitem{thalabard2020butterfly}
S.~Thalabard, J.~Bec, and A.~A. Mailybaev, ``From the butterfly effect to
  spontaneous stochasticity in singular shear flows,'' {\em Communications
  Physics}, vol.~3, no.~1, pp.~1--8, 2020.

\bibitem{drivas2017Alagrangian}
T.~D. Drivas and G.~L. Eyink, ``A lagrangian fluctuation--dissipation relation
  for scalar turbulence. part i. flows with no bounding walls,'' {\em Journal
  of Fluid Mechanics}, vol.~829, pp.~153--189, 2017.

\bibitem{drivas2024statistical}
T.~D. Drivas, A.~A. Mailybaev, and A.~Raibekas, ``Statistical determinism in
  non-lipschitz dynamical systems,'' {\em Ergodic Theory and Dynamical
  Systems}, vol.~44, no.~7, pp.~1856--1884, 2024.

\bibitem{bandak2024spontaneous}
D.~Bandak, A.~Mailybaev, G.~L. Eyink, and N.~Goldenfeld, ``Spontaneous
  stochasticity amplifies even thermal noise to the largest scales of
  turbulence in a few eddy turnover times,'' {\em Phys. Rev. Lett.}, vol.~132,
  p.~104002, 2024.

\bibitem{reynolds1895iv}
O.~Reynolds, ``Iv. on the dynamical theory of incompressible viscous fluids and
  the determination of the criterion,'' {\em Philosophical transactions of the
  royal society of london.(a.)}, no.~186, pp.~123--164, 1895.

\bibitem{taylor1915eddy}
G.~Taylor, ``Eddy motion in the atmosphere,'' {\em Philosophical Transactions
  of the Royal Society of London Series A}, vol.~215, pp.~1--26, 1915.

\bibitem{taylor1922diffusion}
G.~I. Taylor, ``Diffusion by continuous movements,'' {\em Proceedings of the
  London Mathematical Society}, vol.~220, no.~1, pp.~196--212, 1922.

\bibitem{taylor1970some}
G.~Taylor, ``Some early ideas about turbulence,'' {\em Journal of Fluid
  Mechanics}, vol.~41, no.~1, pp.~3--11, 1970.

\bibitem{rayleigh1879stability}
L.~Rayleigh, ``On the stability, or instability, of certain fluid motions,''
  {\em Proceedings of the London Mathematical Society}, vol.~1, no.~1,
  pp.~57--72, 1879.

\bibitem{rayleigh1892viii}
L.~Rayleigh, ``Viii. on the question of the stability of the flow of fluids,''
  {\em The London, Edinburgh, and Dublin Philosophical Magazine and Journal of
  Science}, vol.~34, no.~206, pp.~59--70, 1892.

\bibitem{reynolds1883xxix}
O.~Reynolds, ``Xxix. an experimental investigation of the circumstances which
  determine whether the motion of water shall be direct or sinuous, and of the
  law of resistance in parallel channels,'' {\em Philosophical Transactions of
  the Royal Society of London}, no.~174, pp.~935--982, 1883.

\bibitem{taylor1910conditions}
G.~Taylor, ``The conditions necessary for discontinuous motion in gases,'' {\em
  Proceedings of the Royal Society of London Series A}, vol.~84, no.~571,
  pp.~371--377, 1910.

\bibitem{eyink2022onsager}
G.~L. Eyink, S.~Kumar, and H.~Quan, ``The onsager theory of wall-bounded
  turbulence and taylor's momentum anomaly,'' {\em Philosophical Transactions
  of the Royal Society of London Series A}, vol.~380, no.~2218, p.~20210079,
  2022.

\bibitem{proudman1957expansions}
I.~Proudman and J.~Pearson, ``Expansions at small reynolds numbers for the flow
  past a sphere and a circular cylinder,'' {\em Journal of Fluid Mechanics},
  vol.~2, pp.~237--262, 1957.

\bibitem{stokes1851effect}
G.~Stokes, ``On the effect of the internal friction of fluids on the motion of
  pendulums,'' {\em Transactions of the Cambridge Philosophical Society},
  vol.~9, pp.~8--106, 1851.

\bibitem{veysey2007simple}
J.~Veysey and N.~Goldenfeld, ``Simple viscous flows: From boundary layers to
  the renormalization group,'' {\em Reviews of Modern Physics}, vol.~79, no.~3,
  pp.~883--927, 2007.

\bibitem{goldenfeld1992lectures}
N.~Goldenfeld, {\em Lectures On Phase Transitions And The Renormalization
  Group}.
\newblock Addison-Wesley Reading, MA, 1992.

\bibitem{chen1996renormalization}
L.-Y. Chen, N.~Goldenfeld, and Y.~Oono, ``Renormalization group and singular
  perturbations: Multiple scales, boundary layers, and reductive perturbation
  theory,'' {\em Physical Review E}, vol.~54, no.~1, p.~376, 1996.

\bibitem{Pfenninger1961}
W.~Pfenninger, ``Boundary layer suction experiments with laminar flow at high
  reynolds numbers in the inlet length of a tube by various suction methods,''
  in {\em Boundary Layer and Flow Control} (G.~Lachmann, ed.), pp.~961 -- 980,
  Pergamon Press (Oxford), 1961.
\newblock see appendix starting on pp. 970-980.

\bibitem{linearstability_pipe}
H.~Salwen, F.~W. Cotton, and C.~E. Grosch, ``Linear stability of poiseuille
  flow in a circular pipe,'' {\em Journal of Fluid Mechanics}, vol.~98, no.~2,
  pp.~273--284, 1980.

\bibitem{meseguer2003linearized}
A.~Meseguer and L.~N. Trefethen, ``{Linearized pipe flow to Reynolds number
  $10^7$},'' {\em Journal of Computational Physics}, vol.~186, no.~1,
  pp.~178--197, 2003.

\bibitem{kaneko1984period}
K.~Kaneko, ``Period-doubling of kink-antikink patterns, quasiperiodicity in
  antiferro-like structures and spatial intermittency in coupled logistic
  lattice,'' {\em Progress of Theoretical Physics}, vol.~72, no.~3,
  pp.~480--486, 1984.

\bibitem{janssen1981nonequilibrium}
H.~Janssen, ``On the nonequilibrium phase transition in reaction-diffusion
  systems with an absorbing stationary state,'' {\em Zeitschrift fur Physik B
  Condensed Matter}, vol.~42, no.~2, pp.~151--154, 1981.

\bibitem{grassberger1982phase}
P.~Grassberger, ``On phase transitions in schl{\"o}gl's second model,'' {\em
  Zeitschrift fur Physik B Condensed Matter}, vol.~47, no.~4, pp.~365--374,
  1982.

\bibitem{chate1987transition}
H.~Chate and P.~Manneville, ``Transition to turbulence via spatiotemporal
  intermittency,'' {\em Physical Review Letters}, vol.~58, pp.~112--115, 1987.

\bibitem{shimizu2019exponential}
M.~Shimizu, T.~Kanazawa, and G.~Kawahara, ``Exponential growth of lifetime of
  localized turbulence with its extent in channel flow,'' {\em Fluid Dynamics
  Research}, vol.~51, no.~1, p.~011404, 2019.

\bibitem{mobilia2007}
M.~Mobilia, I.~T. Georgiev, and U.~C. T\"{a}uber, ``Phase transitions and
  spatio-temporal fluctuations in stochastic lattice {L}otka-{V}olterra
  models,'' {\em Journal of Statistical Physics}, vol.~128, no.~1-2,
  pp.~447--483, 2007.

\bibitem{diamond1994}
P.~H. Diamond, Y.-M. Liang, B.~A. Carreras, and P.~W. Terry, ``Self--regulating
  shear flow turbulence: A paradigm for the {L}-{H} transition,'' {\em Phys.
  Rev. Lett.}, vol.~72, pp.~2565--2568, 1994.

\bibitem{estradaPRL2011}
T.~Estrada, C.~Hidalgo, T.~Happel, and P.~H. Diamond, ``Spatiotemporal
  structure of the interaction between turbulence and flows at the {L}-{H}
  transition in a toroidal plasma,'' {\em Phys. Rev. Lett.}, vol.~107,
  p.~245004 (5 pages), 2011.

\bibitem{chen2024mean}
C.~Chen, J.~Tao, and A.~Xu, ``Mean azimuthal flow of puff in pipe flow,'' {\em
  Physics of Fluids}, vol.~36, no.~5, p.~051704, 2024.

\bibitem{khan2024examination}
B.~A. Khan, S.~Arogeti, and A.~Yakhot, ``Examination of the onset and decay of
  turbulence in pipe flow,'' {\em Physical Review Fluids}, vol.~9, no.~9,
  p.~093903, 2024.

\bibitem{waleffe1997self}
F.~Waleffe, ``On a self-sustaining process in shear flows,'' {\em Physics of
  Fluids}, vol.~9, no.~4, pp.~883--900, 1997.

\bibitem{Wang2025stoch}
X.~Wang, H.-Y. Shih, and N.~Goldenfeld, ``Stochastic theory for pattern
  formation and front propagation in transitional pipe turbulence.''
  (Manuscript in preparation), 2025.

\bibitem{ginzburg1958light}
V.~Ginzburg and A.~Levanyuk, ``Light scattering near second-order
  phase-transition and curie points,'' {\em Journal of Physics and Chemistry of
  Solids}, vol.~6, no.~1, pp.~51--58, 1958.

\bibitem{levanyuk1959contribution}
A.~Levanyuk, ``Contribution to the theory of light scattering near the
  second-order phase-transition points,'' {\em Sov. Phys. JETP}, vol.~9, no.~3,
  pp.~571--576, 1959.

\bibitem{ginzburg1961some}
V.~L. Ginzburg, ``Some remarks on phase transitions of the second kind and the
  microscopic theory of ferroelectric materials,'' {\em Fiz. Tverd. Tela},
  vol.~2, pp.~2031--2043, 1960.
\newblock {E}nglish translation published in Soviet Phys. Solid State, 2,
  1824-1834 (1961).

\bibitem{Wang2025SizeCriticalRegion}
X.~Wang and N.~Goldenfeld, ``Size of the critical region for the
  laminar-turbulent transition in quasi-one-dimensional flow geometries.''
  (submitted for publication), 2025.

\bibitem{Shih2026}
H.-Y. Shih and N.~Goldenfeld, ``Super-exponential scaling of puff lifetime near
  the directed percolation transition in pipe turbulence,'' 2026.
\newblock (manuscript in preparation).

\bibitem{bertlmann2000anomalies}
R.~A. Bertlmann, {\em Anomalies in quantum field theory}, vol.~91.
\newblock Oxford university press, 2000.

\bibitem{fujikawa2004path}
K.~Fujikawa and H.~Suzuki, {\em Path integrals and quantum anomalies}.
\newblock No.~122, Oxford University Press, 2004.

\bibitem{mckane2005predator}
A.~McKane and T.~Newman, ``Predator-prey cycles from resonant amplification of
  demographic stochasticity,'' {\em Physical Review Letters}, vol.~94, no.~21,
  p.~218102, 2005.

\bibitem{butler2009robust}
T.~Butler and N.~Goldenfeld, ``Robust ecological pattern formation induced by
  demographic noise,'' {\em Physical Review E—Statistical, Nonlinear, and
  Soft Matter Physics}, vol.~80, no.~3, p.~030902, 2009.

\bibitem{tauber2012population}
U.~C. T{\"a}uber, ``Population oscillations in spatial stochastic
  lotka--volterra models: a field-theoretic perturbational analysis,'' {\em
  Journal of Physics A: Mathematical and Theoretical}, vol.~45, no.~40,
  p.~405002, 2012.

\bibitem{steinberger1949use}
J.~Steinberger, ``On the use of subtraction fields and the lifetimes of some
  types of meson decay,'' {\em Physical Review}, vol.~76, no.~8,
  pp.~1180--1186, 1949.

\bibitem{fukuda1949gamma}
H.~Fukuda and Y.~Miyamoto, ``On the $\gamma$-decay of neutral meson,'' {\em
  Progress of Theoretical Physics}, vol.~4, no.~3, pp.~347--357, 1949.

\bibitem{adler1969axial}
S.~L. Adler, ``Axial-vector vertex in spinor electrodynamics,'' {\em Physical
  Review}, vol.~177, no.~5, pp.~2426--2438, 1969.

\bibitem{bell1969pcac}
J.~Bell and R.~Jackiw, ``A pcac puzzle: $\pi$0→ $\gamma$$\gamma$ in the
  $\sigma$-model,'' {\em Nuovo Cimento A Serie}, vol.~60, no.~1, pp.~47--61,
  1969.

\bibitem{arouca2022quantum}
R.~Arouca, A.~Cappelli, and T.~Hansson, ``Quantum field theory anomalies in
  condensed matter physics,'' {\em arXiv preprint arXiv:2204.02158}, 2022.

\bibitem{mcgreevy2023generalized}
J.~McGreevy, ``Generalized symmetries in condensed matter,'' {\em Annual Review
  of Condensed Matter Physics}, vol.~14, no.~1, pp.~57--82, 2023.

\bibitem{Wang2026anomalies}
X.~Wang and N.~Goldenfeld, ``Stochastic anomalies.'' Manuscript in preparation,
  2026.

\bibitem{onsager1949statistical}
L.~Onsager, ``Statistical hydrodynamics,'' {\em Il Nuovo Cimento (1943-1954)},
  vol.~6, no.~2, pp.~279--287, 1949.

\bibitem{eyink2024onsager}
G.~Eyink, ``Onsager's ‘ideal turbulence’theory,'' {\em Journal of Fluid
  Mechanics}, vol.~988, p.~P1, 2024.

\bibitem{dryden1943review}
H.~L. Dryden, ``A review of the statistical theory of turbulence,'' {\em Q.
  Appl. Math.}, vol.~1, no.~1, pp.~7--42, 1943.

\bibitem{sreenivasan1984scaling}
K.~R. Sreenivasan, ``On the scaling of the turbulence energy dissipation
  rate,'' {\em The Physics of fluids}, vol.~27, no.~5, pp.~1048--1051, 1984.

\bibitem{frisch1995turbulence}
U.~Frisch, {\em Turbulence: the legacy of AN Kolmogorov}.
\newblock Cambridge university press, 1995.

\bibitem{wieselsberger1923ergebnisse}
C.~Wieselsberger, A.~Betz, and L.~Prandtl, {\em Ergebnisse der aerodynamischen
  {V}ersuchsanstalt zu {G}{\"o}ttingen, {II}. {L}ieferung}.
\newblock M\"unchen und Berlin: Druck und Verlag von R. Oldenbourg, 1923.

\bibitem{shoemaker1926resistance}
J.~M. Shoemaker, ``Resistance of a fifteen-centimeter disk,'' Technical Note
  No.~52, National Advisory Committee for Aeronautics, Washington, D.C., 1926.
\newblock
  \url{https://ntrs.nasa.gov/api/citations/19930087660/downloads/19930087660.pdf}.

\bibitem{montgomery1943generalization}
R.~B. Montgomery, ``Generalization for cylinders of {P}randtl's linear
  assumption for mixing length,'' {\em Ann. N. Y. Acad. Sci.}, vol.~44, no.~1,
  pp.~89--103, 1943.

\bibitem{eyink2006onsager}
G.~L. Eyink and K.~R. Sreenivasan, ``Onsager and the theory of hydrodynamic
  turbulence,'' {\em Reviews of modern physics}, vol.~78, no.~1, p.~87, 2006.

\bibitem{nikuradse1932laws}
J.~Nikuradse, ``Gesetzmssigkeiten der turbulenten {S}tromung in glatten
  {R}ohren.'' Verein Deutscher Ingenieure- Forschungsheft 356. Beilage zu
  ``Forschung auf dem Gebiet des Ingenieurdesens'' Ausgabe B, Band 3.
  Translated as "Laws of turbulent flow in smooth pipes,'' National Advisory
  Committee for Aeronautics, Report TT F-10,359, Washington, DC, 1950, 1932.

\bibitem{nikuradse1933laws}
J.~Nikuradse, ``Str\"omungsgesetze in rauhen {R}ohren.'' Verein Deutscher
  Ingenieure - Forschungsheft 361. Beilage zu ``Forschung auf dem Gebiete des
  Ingenieurwesens'' Ausgabe B, Band 4. Translated as "Laws of flow in rough
  pipes,'' National Advisory Committee for Aeronautics, Technical Memorandum
  1292, Washington, DC, 1950, 1933.

\bibitem{cadot1997energy}
O.~Cadot, Y.~Couder, A.~Daerr, S.~Douady, and A.~Tsinober, ``Energy injection
  in closed turbulent flows: {S}tirring through boundary layers versus inertial
  stirring,'' {\em Phys. Rev. E}, vol.~56, no.~1, p.~427, 1997.

\bibitem{pearson2002measurements}
B.~R. Pearson, P.-{\AA}. Krogstad, and W.~van~de Water, ``Measurements of the
  turbulent energy dissipation rate,'' {\em Physics of fluids}, vol.~14, no.~3,
  pp.~1288--1290, 2002.

\bibitem{onsager1945distribution}
L.~Onsager, ``The distribution of energy in turbulence,'' {\em Phys. Rev},
  vol.~68, no.~11-12, p.~286, 1945.

\bibitem{bedrossian2019sufficient}
J.~Bedrossian, M.~Coti~Zelati, S.~Punshon-Smith, and F.~Weber, ``A sufficient
  condition for the kolmogorov 4/5 law for stationary martingale solutions to
  the 3d navier--stokes equations,'' {\em Communications in Mathematical
  Physics}, vol.~367, pp.~1045--1075, 2019.

\bibitem{eyink1994energy}
G.~L. Eyink, ``Energy dissipation without viscosity in ideal hydrodynamics {I}.
  {F}ourier analysis and local energy transfer,'' {\em Physica D}, vol.~78,
  no.~3-4, pp.~222--240, 1994.

\bibitem{delellis2013continuous}
C.~De~Lellis and L.~Sz{\'e}kelyhidi~Jr., ``Continuous dissipative {E}uler flows
  and a conjecture of {O}nsager,'' in {\em European Congress of Mathematics:
  Krak{\'o}w, 2-7 July, 2012} (R.~Lata{\l}a, A.~Ruci{\'n}ski, P.~Strzelecki,
  J.~{\'S}wiatkowski, and D.~Wrzosek, eds.), (Zurich), pp.~13--30, European
  Mathematical Society, 2013.

\bibitem{duchon2000inertial}
J.~Duchon and R.~Robert, ``Inertial energy dissipation for weak solutions of
  incompressible euler and navier-stokes equations,'' {\em Nonlinearity},
  vol.~13, no.~1, p.~249, 2000.

\bibitem{schwinger1951gauge}
J.~Schwinger, ``On gauge invariance and vacuum polarization,'' {\em Phys.
  Rev.}, vol.~82, no.~5, p.~664, 1951.

\bibitem{adler2005anomalies}
S.~L. Adler, ``Anomalies to all orders,'' in {\em 50 years of Yang-Mills
  theory} (G.~Hooft, ed.), pp.~187--228, World Scientific, 2005.

\bibitem{frisch1985singularity}
U.~Frisch and G.~Parisi, ``On the singularity structure of fully developed
  turbulence,'' in {\em Turbulence and Predictability in Geophysical Fluid
  Dynamics and Climate Dynamics} (M.~Ghil, R.~Benzi, and G.~Parisi, eds.),
  vol.~88 of {\em Enrico Fermi International School of Physics Series},
  pp.~84--88, North-Holland, 1985.

\bibitem{lashermes2008comprehensive}
B.~Lashermes, S.~G. Roux, P.~Abry, and S.~Jaffard, ``Comprehensive multifractal
  analysis of turbulent velocity using the wavelet leaders,'' {\em Eur. Phys.
  J. B}, vol.~61, pp.~201--215, 2008.

\bibitem{salmon1988hamiltonian}
R.~Salmon, ``Hamiltonian fluid mechanics,'' {\em Annu. Rev. Fluid Mech.},
  vol.~20, no.~1, pp.~225--256, 1988.

\bibitem{morrison2006hamiltonian}
P.~Morrison, J.~Francoise, and G.~Naber, ``Hamiltonian fluid dynamics,'' {\em
  Encyclopedia of mathematical physics}, vol.~2, pp.~593--600, 2006.

\bibitem{delellis2009euler}
C.~De~Lellis and L.~Sz{\'e}kelyhidi~Jr, ``The euler equations as a differential
  inclusion,'' {\em Annals of mathematics}, pp.~1417--1436, 2009.

\bibitem{delellis2010admissibility}
C.~De~Lellis and L.~Sz{\'e}kelyhidi~Jr, ``On admissibility criteria for weak
  solutions of the euler equations,'' {\em Archive for rational mechanics and
  analysis}, vol.~195, no.~1, pp.~225--260, 2010.

\bibitem{isett2018proof}
P.~Isett, ``A proof of onsager's conjecture,'' {\em Annals of Mathematics},
  vol.~188, no.~3, pp.~871--963, 2018.

\bibitem{buckmaster2019onsager}
T.~Buckmaster, C.~De~Lellis, L.~Sz{\'e}kelyhidi, and V.~Vicol, ``Onsager's
  conjecture for admissible weak solutions,'' {\em Communications on Pure and
  Applied Mathematics}, vol.~72, no.~2, pp.~229--274, 2019.

\bibitem{delellis2019turbulence}
C.~De~Lellis and L.~Sz{\'e}kelyhidi~Jr, ``On turbulence and geometry: from nash
  to onsager,'' {\em Notices of the American Mathematical Society}, vol.~5,
  pp.~677--685, 2019.

\bibitem{daneri2021non}
S.~Daneri, E.~Runa, and L.~Sz{\'e}kelyhidi, ``Non-uniqueness for the euler
  equations up to onsager’s critical exponent,'' {\em Annals of PDE}, vol.~7,
  no.~1, pp.~1--44, 2021.

\bibitem{delellis2023nonuniqueness}
C.~De~Lellis and H.~Kwon, ``On nonuniqueness of h{\"o}lder continuous globally
  dissipative euler flows,'' {\em Analysis \& PDE}, vol.~15, no.~8,
  pp.~2003--2059, 2023.

\bibitem{isett2022nonuniqueness}
P.~Isett, ``Nonuniqueness and existence of continuous, globally dissipative
  euler flows,'' {\em Archive for Rational Mechanics and Analysis}, vol.~244,
  no.~3, pp.~1223--1309, 2022.

\bibitem{giri2023L3}
V.~Giri, H.~Kwon, and M.~Novack, ``The $l^{3}$-based strong onsager theorem,''
  {\em arXiv preprint arXiv:2305.18509}, 2023.

\bibitem{iyer2025turbulence}
K.~P. Iyer, T.~D. Drivas, G.~L. Eyink, and K.~R. Sreenivasan, ``Turbulence
  without walls: Whither the zeroth law of turbulence?,'' {\em Physical Review
  Letters}, vol.~135, no.~13, p.~134001, 2025.

\bibitem{achenbach1972experiments}
E.~Achenbach, ``Experiments on the flow past spheres at very high {R}eynolds
  numbers,'' {\em J. Fluid Mech.}, vol.~54, no.~3, pp.~565--575, 1972.

\bibitem{busse2017reynolds}
A.~Busse, M.~Thakkar, and N.~D. Sandham, ``Reynolds-number dependence of the
  near-wall flow over irregular rough surfaces,'' {\em J. Fluid Mech.},
  vol.~810, pp.~196--224, 2017.

\bibitem{eyink2021josephson}
G.~L. Eyink, ``Josephson-anderson relation and the classical d’alembert
  paradox,'' {\em Physical Review X}, vol.~11, no.~3, p.~031054, 2021.

\bibitem{kumar2024josephson}
S.~Kumar and G.~L. Eyink, ``A josephson--anderson relation for drag in
  classical channel flows with streamwise periodicity: Effects of wall
  roughness,'' {\em Physics of Fluids}, vol.~36, no.~9, 2024.

\bibitem{quan2024onsager}
H.~Quan and G.~L. Eyink, ``Onsager theory of turbulence, the
  josephson--anderson relation, and the d’alembert paradox,'' {\em
  Communications in Mathematical Physics}, vol.~405, no.~11, p.~276, 2024.

\bibitem{bardos2018onsager}
C.~Bardos and E.~S. Titi, ``Onsager’s conjecture for the incompressible euler
  equations in bounded domains,'' {\em Archive for Rational Mechanics and
  Analysis}, vol.~228, no.~1, pp.~197--207, 2018.

\bibitem{quan2025inertial}
H.~Quan and G.~L. Eyink, ``Inertial momentum dissipation for viscosity
  solutions of euler equations: External flow around a smooth body,'' {\em
  Nonlinearity}, vol.~38, no.~10, p.~105019, 2025.

\bibitem{eyink2025weak}
G.~Eyink and H.~Quan, ``Weak-strong uniqueness and extreme wall events at high
  reynolds number,'' {\em Physical Review Fluids}, vol.~10, no.~6, p.~064610,
  2025.

\bibitem{barenblatt1972self}
G.~I. Barenblatt and Y.~B. Zel'dovich, ``Self-similar solutions as intermediate
  asymptotics,'' {\em Annual Review of Fluid Mechanics}, vol.~4, pp.~285--312,
  1972.

\bibitem{barenblatt1996scaling}
G.~I. Barenblatt, {\em Scaling, self-similarity, and intermediate asymptotics}.
\newblock Cambridge University Press, 1996.

\bibitem{adler1969absence}
S.~L. Adler and W.~A. Bardeen, ``Absence of higher-order corrections in the
  anomalous axial-vector divergence equation,'' {\em Physical Review},
  vol.~182, no.~5, p.~1517, 1969.

\bibitem{zee1972axial}
A.~Zee, ``Axial-vector anomalies and the scaling property of field theory,''
  {\em Physical Review Letters}, vol.~29, no.~17, p.~1198, 1972.

\bibitem{BLAS13}
H.~Blasius, ``Das ahnlichkeitsgesetz bei reibungsvorg\"angen in
  fl\"ussigkeiten,'' {\em Forsch. Arb. Ing. Wes. No. 134, Berlin}, 1913.

\bibitem{STRI23}
A.~Strickler, ``Beitrage zur frage der geschwindigkeitsformel und der
  rauhigkeitszahlen fur strome, kanale und geschlossene leitungen,'' {\em
  Mitteilungen des Eidgenossischen Amtes fur Wasserwirtschaft 16, Bern,
  Switzerland ~Translated as \lq\lq Contributions to the question of a velocity
  formula and roughness data for streams, channels and closed pipelines." by T.
  Roesgan and W. R. Brownie, Translation T-10, W. M. Keck Lab of Hydraulics and
  Water Resources, Calif. Inst. Tech., Pasadena, Calif. January 1981}, 1923.

\bibitem{goldenfeld2006roughness}
N.~Goldenfeld, ``Roughness-induced critical phenomena in a turbulent flow,''
  {\em Physical review letters}, vol.~96, no.~4, p.~044503, 2006.

\bibitem{mehrafarin2008intermittency}
M.~Mehrafarin and N.~Pourtolami, ``Intermittency and rough-pipe turbulence,''
  {\em Physical Review E}, vol.~77, no.~5, p.~055304, 2008.

\bibitem{gioia2006turbulent}
G.~Gioia and P.~Chakraborty, ``Turbulent friction in rough pipes and the energy
  spectrum of the phenomenological theory,'' {\em Physical Review Letters},
  vol.~96, no.~4, p.~044502, 2006.

\bibitem{guttenberg2009friction}
N.~Guttenberg and N.~Goldenfeld, ``Friction factor of two-dimensional
  rough-boundary turbulent soap film flows,'' {\em Physical Review E}, vol.~79,
  no.~6, p.~065306, 2009.

\bibitem{tran2010macroscopic}
T.~Tran, P.~Chakraborty, N.~Guttenberg, A.~Prescott, H.~Kellay, W.~Goldburg,
  N.~Goldenfeld, and G.~Gioia, ``Macroscopic effects of the spectral structure
  in turbulent flows,'' {\em Nature Physics}, vol.~6, no.~6, pp.~438--441,
  2010.

\bibitem{kellay2012testing}
H.~Kellay, T.~Tran, W.~Goldburg, N.~Goldenfeld, G.~Gioia, and P.~Chakraborty,
  ``Testing a missing spectral link in turbulence,'' {\em Physical Review
  Letters}, vol.~109, no.~25, p.~254502, 2012.

\bibitem{vilquin2021asymptotic}
A.~Vilquin, J.~Jagielka, S.~Djambov, H.~Herouard, P.~Fischer, C.-H. Bruneau,
  P.~Chakraborty, G.~Gioia, and H.~Kellay, ``Asymptotic turbulent friction in
  2d rough-walled flows,'' {\em Science Advances}, vol.~7, no.~5, p.~eabc6234,
  2021.

\bibitem{wyngaard1972some}
J.~C. Wyngaard and Y.~H. Pao, ``Some measurements of the fine structure of
  large {R}eynolds number turbulence,'' in {\em Statistical Models and
  Turbulence} (M.~Rosenblatt and C.~Van~Atta, eds.), (Berlin, Heidelberg),
  pp.~384--401, Springer Berlin Heidelberg, 1972.

\bibitem{eyink1994analogies}
G.~Eyink and N.~Goldenfeld, ``Analogies between scaling in turbulence, field
  theory, and critical phenomena,'' {\em Physical Review E}, vol.~50, no.~6,
  p.~4679, 1994.

\bibitem{falkovich2001particles}
G.~Falkovich, K.~Gaw\c{e}dzki, and M.~Vergassola, ``Particles and fields in
  fluid turbulence,'' {\em Reviews of modern Physics}, vol.~73, no.~4, p.~913,
  2001.

\bibitem{kraichnan1968small}
R.~H. Kraichnan, ``Small-scale structure of a scalar field convected by
  turbulence,'' {\em The Physics of Fluids}, vol.~11, no.~5, pp.~945--953,
  1968.

\bibitem{kupiainen2007scaling}
A.~Kupiainen and P.~Muratore-Ginanneschi, ``Scaling, renormalization and
  statistical conservation laws in the kraichnan model of turbulent
  advection,'' {\em Journal of Statistical Physics}, vol.~126, no.~3,
  pp.~669--724, 2007.

\bibitem{eyink1993lagrangian}
G.~L. Eyink, ``Lagrangian field theory, multifractals, and universal scaling in
  turbulence,'' {\em Physics Letters A}, vol.~172, no.~5, pp.~355--360, 1993.

\bibitem{buaria2025intermittency}
D.~Buaria, ``Intermittency of longitudinal and transverse velocity gradients in
  turbulence and their relation to inertial range exponents,'' in {\em Division
  of Fluid Dynamics Annual Meeting 2025}, APS, 2025.

\bibitem{rosenhaus2024wave}
V.~Rosenhaus and M.~Smolkin, ``Wave turbulence and the kinetic equation beyond
  leading order,'' {\em Physical Review E}, vol.~109, no.~6, p.~064127, 2024.

\bibitem{rosenhaus2024interaction}
V.~Rosenhaus and G.~Falkovich, ``Interaction renormalization and validity of
  kinetic equations for turbulent states,'' {\em Physical Review Letters},
  vol.~133, no.~24, p.~244002, 2024.

\bibitem{rosenhaus2025weak}
V.~Rosenhaus and G.~Falkovich, ``Weak and strong turbulence in self-focusing
  and defocusing media,'' {\em arXiv preprint arXiv:2501.12451}, 2025.

\bibitem{donzis2005scalar}
D.~Donzis, K.~Sreenivasan, and P.~K. Yeung, ``Scalar dissipation rate and
  dissipative anomaly in isotropic turbulence,'' {\em Journal of Fluid
  Mechanics}, vol.~532, pp.~199--216, 2005.

\bibitem{obukhov1949structure}
A.~M. Obukhov {\em et~al.}, ``Structure of the temperature field in a turbulent
  flow,'' {\em Izv. Akad. Nauk SSSR, Ser. Geogr. Geofiz}, vol.~13, no.~1,
  pp.~58--69, 1949.

\bibitem{corrsin1951spectrum}
S.~Corrsin, ``On the spectrum of isotropic temperature fluctuations in an
  isotropic turbulence,'' {\em Journal of Applied Physics}, vol.~22, no.~4,
  pp.~469--473, 1951.

\bibitem{gawedzki1998intermittency}
K.~Gaw\c{e}dzki, ``Intermittency of passive advection,'' in {\em Advances in
  Turbulence VII: Proceedings of the Seventh European Turbulence Conference,
  held in Saint-Jean Cap Ferrat, France, 30 June -- 3 July 1998 / Actes de la
  Septi{\`e}me Conf{\'e}rence Europ{\'e}enne de Turbulence, tenue {\`a}
  Saint-Jean Cap Ferrat, France, 30 Juin -- 3 Juillet 1998} (U.~Frisch, ed.),
  pp.~493--502, Kluwer Academic Publishers, 1998.

\bibitem{hartman1982ordinary}
P.~Hartman, {\em Ordinary Differential Equations: Second Edition}, vol.~38 of
  {\em Classics in Applied Mathematics}.
\newblock Philadelphia, USA: Society for Industrial and Applied Mathematics,
  1982.

\bibitem{agarwal1993uniqueness}
R.~Agarwal, R.~Agarwal, and V.~Lakshmikantham, {\em Uniqueness and
  Nonuniqueness Criteria for Ordinary Differential Equations}, vol.~6 of {\em
  Series in real analysis}.
\newblock Singapore: World Scientific Publishing, 1993.

\bibitem{richardson1926atmospheric}
L.~F. Richardson, ``Atmospheric diffusion shown on a distance-neighbour
  graph,'' {\em Proceedings of the Royal Society of London. Series A},
  vol.~110, no.~756, pp.~709--737, 1926.

\bibitem{chaves2003lagrangian}
M.~Chaves, K.~Gawedzki, P.~Horvai, A.~Kupiainen, and M.~Vergassola,
  ``Lagrangian dispersion in gaussian self-similar velocity ensembles,'' {\em
  Journal of statistical physics}, vol.~113, no.~5, pp.~643--692, 2003.

\bibitem{peixoto2023spontaneous}
A.~L. Peixoto~Considera and S.~Thalabard, ``Spontaneous stochasticity in the
  presence of intermittency,'' {\em Physical Review Letters}, vol.~131, no.~6,
  p.~064001, 2023.

\bibitem{armstrong2025anomalous}
S.~Armstrong and V.~Vicol, ``Anomalous diffusion by fractal homogenization,''
  {\em Annals of PDE}, vol.~11, no.~1, p.~2, 2025.

\bibitem{ruffenach2025spontaneous}
W.~Ruffenach, E.~Simonnet, and N.~Valade, ``Spontaneous stochasticity and the
  armstrong-vicol passive scalar,'' {\em arXiv preprint arXiv:2504.15795},
  2025.

\bibitem{burczak2023anomalous}
J.~Burczak, L.~Sz{\'e}kelyhidi~Jr, and B.~Wu, ``Anomalous dissipation and euler
  flows,'' {\em arXiv preprint arXiv:2310.02934}, 2023.

\bibitem{drivas2017Blagrangian}
T.~D. Drivas and G.~L. Eyink, ``A lagrangian fluctuation--dissipation relation
  for scalar turbulence. part ii. wall-bounded flows,'' {\em Journal of Fluid
  Mechanics}, vol.~829, pp.~236--279, 2017.

\bibitem{eyink2011stochastic}
G.~L. Eyink, ``Stochastic flux freezing and magnetic dynamo,'' {\em Physical
  Review E—Statistical, Nonlinear, and Soft Matter Physics}, vol.~83, no.~5,
  p.~056405, 2011.

\bibitem{buaria2016lagrangian}
D.~Buaria, P.-K. Yeung, and B.~L. Sawford, ``A lagrangian study of turbulent
  mixing: forward and backward dispersion of molecular trajectories in
  isotropic turbulence,'' {\em Journal of Fluid Mechanics}, vol.~799,
  pp.~352--382, 2016.

\bibitem{bitane2013geometry}
R.~Bitane, H.~Homann, and J.~Bec, ``Geometry and violent events in turbulent
  pair dispersion,'' {\em Journal of Turbulence}, vol.~14, no.~2, pp.~23--45,
  2013.

\bibitem{buaria2015characteristics}
D.~Buaria, B.~L. Sawford, and P.-K. Yeung, ``Characteristics of backward and
  forward two-particle relative dispersion in turbulence at different reynolds
  numbers,'' {\em Physics of Fluids}, vol.~27, no.~10, 2015.

\bibitem{bourgoin2006role}
M.~Bourgoin, N.~T. Ouellette, H.~Xu, J.~Berg, and E.~Bodenschatz, ``The role of
  pair dispersion in turbulent flow,'' {\em Science}, vol.~311, no.~5762,
  pp.~835--838, 2006.

\bibitem{ouellette2006experimental}
N.~T. Ouellette, H.~Xu, M.~Bourgoin, and E.~Bodenschatz, ``An experimental
  study of turbulent relative dispersion models,'' {\em New Journal of
  Physics}, vol.~8, no.~6, p.~109, 2006.

\bibitem{tan2022universality}
S.~Tan and R.~Ni, ``Universality and intermittency of pair dispersion in
  turbulence,'' {\em Physical Review Letters}, vol.~128, no.~11, p.~114502,
  2022.

\bibitem{goldenfeld1991asymptotics}
N.~Goldenfeld, O.~Martin, and Y.~Oono, ``Asymptotics of partial differential
  equations and the renormalisation group,'' in {\em Asymptotics Beyond All
  Orders}, pp.~375--383, Springer, 1991.

\bibitem{chen1994selection}
L.-Y. Chen, N.~Goldenfeld, Y.~Oono, and G.~C. Paquette, ``Selection, stability
  and renormalization,'' {\em Physica A: Statistical Mechanics and its
  Applications}, vol.~204, no.~1-4, pp.~111--133, 1994.

\bibitem{bricmont1995renormalizing}
J.~Bricmont and A.~Kupiainen, ``Renormalizing partial differential equations,''
  in {\em Constructive Physics Results in Field Theory, Statistical Mechanics
  and Condensed Matter Physics}, pp.~83--115, Springer, 1995.

\bibitem{bricmont1999stability}
J.~Bricmont, A.~Kupiainen, and J.~Taskinen, ``Stability of {C}ahn-{H}illiard
  fronts,'' {\em Communications on Pure and Applied Mathematics}, vol.~52,
  no.~7, pp.~839--871, 1999.

\bibitem{eyink2020renormalization}
G.~L. Eyink and D.~Bandak, ``Renormalization group approach to spontaneous
  stochasticity,'' {\em Physical Review Research}, vol.~2, no.~4, p.~043161,
  2020.

\bibitem{eyink2014mathematical}
G.~L. Eyink, ``Mathematical {A}nalysis of {T}urbulence {III}.'' Mathematics of
  Turbulence Tutorials, September 9 - 12, 2014. Part of the Long Program,
  ``Mathematics of Turbulence'', Institute for Pure and Applied Mathematics,
  UCLA, Los Angeles, CA;
  \url{https://www.ipam.ucla.edu/programs/workshops/mathematics-of-turbulence-tutorials/?tab=schedule},
  2014.

\bibitem{leith1972predictability}
C.~Leith and R.~Kraichnan, ``Predictability of turbulent flows,'' {\em Journal
  of Atmospheric Sciences}, vol.~29, no.~6, pp.~1041--1058, 1972.

\bibitem{boffetta2001predictability}
G.~Boffetta and S.~Musacchio, ``Predictability of the inverse energy cascade in
  2d turbulence,'' {\em Physics of Fluids}, vol.~13, no.~4, pp.~1060--1062,
  2001.

\bibitem{boffetta2017chaos}
G.~Boffetta and S.~Musacchio, ``Chaos and predictability of
  homogeneous-isotropic turbulence,'' {\em Physical review letters}, vol.~119,
  no.~5, p.~054102, 2017.

\bibitem{lorenz1963deterministic}
E.~N. Lorenz, ``Deterministic nonperiodic flow,'' {\em Journal of the
  atmospheric sciences}, vol.~20, no.~2, pp.~130--141, 1963.

\bibitem{palmer2024real}
T.~Palmer, ``The real butterfly effect and maggoty apples,'' {\em Physics
  Today}, vol.~77, no.~5, pp.~30--35, 2024.

\bibitem{drivas2021life}
T.~D. Drivas and A.~A. Mailybaev, ``‘life after death’in ordinary
  differential equations with a non-lipschitz singularity,'' {\em
  Nonlinearity}, vol.~34, no.~4, p.~2296, 2021.

\bibitem{mailybaev2023spontaneously}
A.~A. Mailybaev and A.~Raibekas, ``Spontaneously stochastic arnold’s cat,''
  {\em Arnold Mathematical Journal}, vol.~9, no.~3, pp.~339--357, 2023.

\bibitem{yamada1998asymptotic}
M.~Yamada and K.~Ohkitani, ``Asymptotic formulas for the lyapunov spectrum of
  fully developed shell model turbulence,'' {\em Physical Review E}, vol.~57,
  no.~6, p.~R6257, 1998.

\bibitem{mailybaev2017toward}
A.~A. Mailybaev, ``Toward analytic theory of the rayleigh--taylor instability:
  lessons from a toy model,'' {\em Nonlinearity}, vol.~30, no.~6, p.~2466,
  2017.

\bibitem{biferale2018rayleigh}
L.~Biferale, G.~Boffetta, A.~A. Mailybaev, and A.~Scagliarini,
  ``Rayleigh-taylor turbulence with singular nonuniform initial conditions,''
  {\em Physical Review Fluids}, vol.~3, no.~9, p.~092601, 2018.

\bibitem{szekelyhidi2011weak}
L.~Sz{\'e}kelyhidi~Jr, ``Weak solutions to the incompressible euler equations
  with vortex sheet initial data,'' {\em Comptes Rendus Mathematique},
  vol.~349, no.~19-20, pp.~1063--1066, 2011.

\bibitem{mengual2023dissipative}
F.~Mengual and L.~Sz{\'e}kelyhidi~Jr, ``Dissipative euler flows for vortex
  sheet initial data without distinguished sign,'' {\em Communications on Pure
  and Applied Mathematics}, vol.~76, no.~1, pp.~163--221, 2023.

\bibitem{gebhard2021new}
B.~Gebhard, J.~J. Kolumb{\'a}n, and L.~Sz{\'e}kelyhidi~Jr, ``A new approach to
  the rayleigh--taylor instability,'' {\em Archive for rational mechanics and
  analysis}, vol.~241, no.~3, pp.~1243--1280, 2021.

\bibitem{gebhard2025rayleigh}
B.~Gebhard and J.~J. Kolumb{\'a}n, ``The rayleigh-taylor instability with local
  energy dissipation,'' {\em arXiv preprint arXiv:2505.03278}, 2025.

\bibitem{betchov1957fine}
R.~Betchov, ``On the fine structure of turbulent flows,'' {\em Journal of Fluid
  Mechanics}, vol.~3, no.~2, pp.~205--216, 1957.

\bibitem{betchov1961thermal}
R.~Betchov, ``Thermal agitation and turbulence,'' in {\em Rarefied Gas
  Dynamics} (L.~Talbot, ed.), p.~307–321, New York: Academic Press, 1961.
\newblock Proceedings of the Second International Symposium on Rarefied Gas
  Dynamics, held at the University of California, Berkeley, CA, 1960.

\bibitem{bandak2022dissipation}
D.~Bandak, N.~Goldenfeld, A.~A. Mailybaev, and G.~Eyink, ``Dissipation-range
  fluid turbulence and thermal noise,'' {\em Physical Review E}, vol.~105,
  no.~6, p.~065113, 2022.

\bibitem{bell_nonaka_garcia_eyink_2022}
J.~B. Bell, A.~Nonaka, A.~L. Garcia, and G.~Eyink, ``Thermal fluctuations in
  the dissipation range of homogeneous isotropic turbulence,'' {\em Journal of
  Fluid Mechanics}, vol.~939, p.~A12, 2022.

\bibitem{forster1977large}
D.~Forster, D.~R. Nelson, and M.~J. Stephen, ``Large-distance and long-time
  properties of a randomly stirred fluid,'' {\em Physical Review A}, vol.~16,
  no.~2, p.~732, 1977.

\bibitem{donev2014low}
A.~Donev, A.~Nonaka, Y.~Sun, T.~Fai, A.~Garcia, and J.~Bell, ``Low mach number
  fluctuating hydrodynamics of diffusively mixing fluids,'' {\em Communications
  in Applied Mathematics and Computational Science}, vol.~9, no.~1,
  pp.~47--105, 2014.

\bibitem{ortiz2025spontaneous}
E.~Ortiz, C.~S. Campolina, and A.~A. Mailybaev, ``Spontaneous stochasticity in
  the fluctuating navier-stokes equations on a logarithmic lattice,'' {\em
  arXiv preprint arXiv:2507.03196}, 2025.

\bibitem{ruelle1979microscopic}
D.~Ruelle, ``Microscopic fluctuations and turbulence,'' {\em Physics Letters
  A}, vol.~72, no.~2, pp.~81--82, 1979.

\bibitem{eyink2025space}
G.~L. Eyink and L.~Peng, ``Space-time statistical solutions of the
  incompressible euler equations and landau-lifshitz fluctuating
  hydrodynamics,'' {\em Nonlinearity}, vol.~38, no.~8, p.~085011, 2025.

\bibitem{mailybaev2023spontaneous}
A.~A. Mailybaev and A.~Raibekas, ``Spontaneous stochasticity and
  renormalization group in discrete multi-scale dynamics,'' {\em Commun. Math.
  Phys.}, vol.~401, pp.~2643--2671, 2023.

\bibitem{mailybaev2025rgA}
A.~A. Mailybaev, ``Rg approach to the inviscid limit for shell models of
  turbulence,'' {\em Nonlinearity}, vol.~38, no.~8, p.~085010, 2025.

\bibitem{mailybaev2025rgB}
A.~A. Mailybaev, ``Rg analysis of spontaneous stochasticity on a fractal
  lattice: stability and bifurcations,'' {\em Journal of Statistical Physics},
  vol.~192, no.~3, pp.~1--22, 2025.

\bibitem{mailybaev2025rgC}
A.~A. Mailybaev, ``Rg theory of spontaneous stochasticity for sabra model of
  turbulence,'' {\em arXiv preprint arXiv:2510.01204}, 2025.

\bibitem{peng2025spontaneous}
L.~Peng, {\em Spontaneous Stochasticity and the Space-Time Statistical
  Solutions of Landau–Lifshitz Hydrodynamics}.
\newblock PhD thesis, The Johns Hopkins University, 2025.

\bibitem{parisi2023nobel}
G.~Parisi, ``Nobel lecture: Multiple equilibria,'' {\em Reviews of Modern
  Physics}, vol.~95, no.~3, p.~030501, 2023.

\bibitem{lototskii2004passive}
S.~V. Lototskii and B.~L. Rozovskii, ``Passive scalar equation in a turbulent
  incompressible gaussian velocity field,'' {\em Russian Mathematical Surveys},
  vol.~59, no.~2, p.~297, 2004.

\bibitem{eyink2018lagrangian}
G.~L. Eyink and T.~D. Drivas, ``A lagrangian fluctuation--dissipation relation
  for scalar turbulence. part iii. turbulent rayleigh--b{\'e}nard convection,''
  {\em Journal of Fluid Mechanics}, vol.~836, pp.~560--598, 2018.

\bibitem{lazarian1999reconnection}
A.~Lazarian and E.~T. Vishniac, ``Reconnection in a weakly stochastic field,''
  {\em The Astrophysical Journal}, vol.~517, no.~2, p.~700, 1999.

\bibitem{eyink2011fast}
G.~L. Eyink, A.~Lazarian, and E.~T. Vishniac, ``Fast magnetic reconnection and
  spontaneous stochasticity,'' {\em The Astrophysical Journal}, vol.~743,
  no.~1, p.~51, 2011.

\bibitem{eyink2013flux}
G.~Eyink, E.~Vishniac, C.~Lalescu, H.~Aluie, K.~Kanov, K.~B{\"u}rger, R.~Burns,
  C.~Meneveau, and A.~Szalay, ``Flux-freezing breakdown in high-conductivity
  magnetohydrodynamic turbulence,'' {\em Nature}, vol.~497, no.~7450,
  pp.~466--469, 2013.

\bibitem{tang2025turbulent}
Y.~Tang, S.~Inui, Y.~Xing, Y.~Qi, and W.~Guo, ``Turbulent diffusion and
  dispersion in a superfluid,'' {\em arXiv preprint arXiv:2504.00353}, 2025.

\bibitem{krommes1994role}
J.~A. Krommes and G.~Hu, ``The role of dissipation in the theory and
  simulations of homogeneous plasma turbulence, and resolution of the entropy
  paradox,'' {\em Physics of plasmas}, vol.~1, no.~10, pp.~3211--3238, 1994.

\bibitem{schekochihin2009astrophysical}
A.~Schekochihin, S.~C. Cowley, W.~Dorland, G.~Hammett, G.~G. Howes,
  E.~Quataert, and T.~Tatsuno, ``Astrophysical gyrokinetics: kinetic and fluid
  turbulent cascades in magnetized weakly collisional plasmas,'' {\em The
  Astrophysical Journal Supplement Series}, vol.~182, no.~1, p.~310, 2009.

\bibitem{bardos2020onsager}
C.~Bardos, N.~Besse, and T.~T. Nguyen, ``Onsager-type conjecture and
  renormalized solutions for the relativistic vlasov--maxwell system,'' {\em
  Quarterly of Applied Mathematics}, vol.~78, no.~2, pp.~193--217, 2020.

\bibitem{athanassoulis2012strong}
A.~Athanassoulis and T.~Paul, ``Strong and weak semiclassical limit for some
  rough hamiltonians,'' {\em Mathematical Models and Methods in Applied
  Sciences}, vol.~22, no.~12, p.~1250038, 2012.

\bibitem{eyink2015quantum}
G.~L. Eyink and T.~D. Drivas, ``Quantum spontaneous stochasticity,'' {\em arXiv
  preprint arXiv:1509.04941}, 2015.

\bibitem{hasselmann1976stochastic}
K.~Hasselmann, ``Stochastic climate models part i. theory,'' {\em tellus},
  vol.~28, no.~6, pp.~473--485, 1976.

\bibitem{rose1977eddy}
H.~A. Rose, ``Eddy diffusivity, eddy noise and subgrid-scale modelling,'' {\em
  Journal of Fluid Mechanics}, vol.~81, no.~4, pp.~719--734, 1977.

\bibitem{leith1990stochastic}
C.~Leith, ``Stochastic backscatter in a subgrid-scale model: Plane shear mixing
  layer,'' {\em Physics of Fluids A: Fluid Dynamics}, vol.~2, no.~3,
  pp.~297--299, 1990.

\bibitem{eyink1996turbulence}
G.~L. Eyink, ``Turbulence noise,'' {\em Journal of statistical physics},
  vol.~83, no.~5, pp.~955--1019, 1996.

\bibitem{palmer2019stochastic}
T.~Palmer, ``Stochastic weather and climate models,'' {\em Nature Reviews
  Physics}, vol.~1, no.~7, pp.~463--471, 2019.

\bibitem{genel2019quantification}
S.~Genel, G.~L. Bryan, V.~Springel, L.~Hernquist, D.~Nelson, A.~Pillepich,
  R.~Weinberger, R.~Pakmor, F.~Marinacci, and M.~Vogelsberger, ``A
  quantification of the butterfly effect in cosmological simulations and
  implications for galaxy scaling relations,'' {\em The Astrophysical Journal},
  vol.~871, no.~1, p.~21, 2019.

\bibitem{neyrinck2022boundaries}
M.~Neyrinck, S.~Genel, and J.~St{\"u}cker, ``Boundaries of chaos and
  determinism in the cosmos,'' {\em arXiv preprint arXiv:2206.10666}, 2022.

\end{thebibliography}

\end{document}